\documentclass[12pt]{article}
\pdfoutput=1

\usepackage{putex}
\usepackage{graphicx}
\usepackage{caption}
\usepackage{amsmath}
\usepackage{array}
\usepackage{subcaption}
\usepackage{enumerate}
\usepackage{cite}
\usepackage{slashed}
\usepackage{bm}
\usepackage{bbm}
\usepackage[utf8]{inputenc}
\usepackage[
      colorlinks=true,
      linkcolor=blue,
      urlcolor=blue,
      filecolor=black,
      citecolor=red,
      ]{hyperref}
\usepackage{braket}
\usepackage{todonotes}
\usepackage{tikz}
\usetikzlibrary{decorations.markings}
\usepackage{tikz-feynman}
\usepackage{caption}
\newcommand{\es}[2] {\begin{equation} \label{#1} \begin{split} #2 \end{split} \end{equation}}

\newcommand{\Z}{\mathbb{Z}}

\newcommand{\oeis}[1]{\href{http://oeis.org/#1}{#1}}

\newcommand{\rep}[1]{\bm{#1}}

\numberwithin{equation}{section}

\definecolor{purple}{rgb}{0.6,0.1,1}

\def\hc{{\rm h.c.}}
\DeclareMathOperator{\Tr}{Tr}
\DeclareMathOperator{\SU}{SU}
\DeclareMathOperator{\SO}{SO}

\DeclareMathOperator{\adj}{\textbf{adj}}
\DeclareMathOperator{\poly}{poly}
\DeclareMathOperator{\PE}{PE}
\DeclareMathOperator{\rk}{rk}

\begin{document}

\preprint{PUPT-2635}

\institution{PU}{Joseph Henry Laboratories, Princeton University, Princeton, NJ 08544, USA}
\institution{PCTS}{Princeton Center for Theoretical Science, Princeton University, Princeton, NJ 08544}

\title{Adjoint Majorana QCD$_2$ at Finite $N$}

\authors{Ross Dempsey,\worksat{\PU} Igor R.~Klebanov,\worksat{\PU, \PCTS} Loki L. Lin,\worksat{\PU} and Silviu S.~Pufu\worksat{\PU}}

\abstract{The mass spectrum of $1+1$-dimensional $\SU(N)$ gauge theory coupled to a Majorana fermion in the adjoint representation has been studied in the large $N$ limit using Light-Cone Quantization. Here we extend this approach to theories with small values of $N$, exhibiting explicit results for $N=2, 3$, and $4$. In the context of
Discretized Light-Cone Quantization, we develop a procedure based on the Cayley-Hamilton theorem for determining which states of the large $N$ theory become null at finite $N$.
For the low-lying bound states, we find that the squared masses divided by $g^2 N$, where $g$ is the gauge coupling, have very weak dependence on $N$. The coefficients of the $1/N^2$ corrections to their large $N$ values
are surprisingly small.  When the adjoint fermion is massless, we observe exact degeneracies that we explain in terms of a Kac-Moody algebra construction and charge conjugation symmetry.  When the squared mass of the adjoint fermion is tuned to $g^2 N / \pi$, we find evidence that the spectrum exhibits boson-fermion degeneracies, in agreement with the supersymmetry of the model at any value of $N$.
}
\date{October 2022}

\maketitle

\tableofcontents

\section{Introduction}

The non-perturbative dynamics of quantum gauge theories is an important research area, with applications to several fields of physics including the strong nuclear interactions and condensed matter systems. The non-abelian gauge theories in 1+1 dimensions, which have served as simplified models for Quantum Chromodynamics (QCD), provide a nice playground where various methods can be tested. They are also interesting in their own right and may have connections to condensed matter and cold atom physics. A well-known such 2D model is the $\SU(N)$ gauge gauge theory coupled to a fundamental Dirac fermion of mass $m$, often called QCD$_2$. 't Hooft \cite{tHooft:1974pnl} used Light-Cone Quantization (LCQ) to show that, in the large $N$ limit, the meson spectrum of this model is calculable and
consists of a single ``Regge trajectory." Therefore, this model exhibits quark confinement. The LCQ has also been 
applied to the QCD$_2$ models with small numbers of colors $N$ \cite{Hornbostel:1988fb,Hornbostel:1988ne,Anand:2021qnd}. This is more complicated than the large $N$ limit, since the states can no longer be truncated to a single quark-antiquark pair. Nevertheless, numerical diagonalizations of the Light-Cone Hamiltonian have produced quite precise results. Thus, the LCQ approach to the bound state spectrum is sometimes more efficient than the Lattice Gauge Theory \cite{Hamer:1981yq,Banuls:2017ena} or bosonization techniques \cite{Frishman:1992mr}. When these different numerical approaches can be compared, they appear to be in good agreement with each other \cite{Hornbostel:1988ne,Anand:2021qnd}. A remarkable feature of QCD$_2$ with a massless quark is that the spectrum contains a massless baryon in addition to a massless meson \cite{Steinhardt:1980ry,Amati:1981fn}.

A key reason for the simplification in two spacetime dimensions is that the gauge field is not dynamical. This is a sharp distinction from the four-dimensional case, in which propagating gluon degrees of freedom are physically important. It is thus interesting to consider generalizations of the 't Hooft model containing fields in the adjoint 
representation of $\SU(N)$ \cite{Dalley:1992yy}. The theory then has propagating adjoint degrees of freedom that lead to interesting dynamics and the presence of different topological sectors \cite{Witten:1978ka,Smilga:1994hc,Lenz:1994du,Cherman:2019hbq,Komargodski:2020mxz}. In the large $N$ limit, application of Light-Cone Quantization leads to truncation to the single-trace states. Since in the adjoint case these closed string-like states can consist of arbitrarily large numbers of adjoint quanta, the numerical solution of the Light-Cone Schr\"odinger equation is necessarily much more complicated than for the 't Hooft model. Nevertheless, good numerical results for the spectra have been obtained using Discretized Light-Cone Quantization (DLCQ) \cite{Pauli:1985ps} as well as Conformal Truncation approaches \cite{Katz:2013qua,Anand:2020gnn}.  

A particularly simple 2D model with adjoint matter is $\SU(N)$ gauge theory coupled to an adjoint
Majorana fermion \cite{Dalley:1992yy,Kutasov:1993gq,Bhanot:1993xp,Demeterfi:1993rs}. The Light-Cone Quantization of this model is not afflicted by the fermion-doubling problems seen on the lattice, and very good convergence of the large $N$ bound state spectra in DLCQ has been observed \cite{Bhanot:1993xp,dempsey_exact_2021}.
When the adjoint mass $m_\text{adj}$ is taken to 0, all the ``gluinoballs" stay massive, which follows from the vanishing of the IR central charge \cite{Kutasov:1993gq,Bhanot:1993xp}; this gauged Majorana model is thus an example of a gapped topological phase. A remarkable feature of the $m_\text{adj}\rightarrow 0$ limit is that the string tension divided by $g^2 N$ is renormalized from order $1$ at short distances to zero at long distances. Thus, the theory with a massless adjoint fermion is not confining \cite{Gross:1995bp,Gross:1997mx,Komargodski:2020mxz,dempsey_exact_2021,delmastro_infrared_2021}. In the DLCQ approach with antiperiodic boundary conditions, the lack of confinement manifests itself in certain exact degeneracies observed even at a finite value of the cut-off $K$ \cite{Gross:1997mx,dempsey_exact_2021}. These degeneracies arise due to the Kac-Moody symmetry present in the DLCQ formulation of the model \cite{Kutasov:1994xq, dempsey_exact_2021}.

We can further improve upon the analogy with physical QCD by passing from the large $N$ limit to finite $N$ and considering a 2D Yang-Mills theory with a low-rank $\SU(N)$ gauge group coupled to an adjoint Majorana fermion of mass $m_\text{adj}$. This theory was previously studied by Antonuccio and Pinsky in~\cite{Antonuccio:1998uz}, who used DLCQ to numerically estimate the masses of the lowest fermion and boson states. However, their method is not straightforwardly extensible to extracting the entire spectrum, for group-theoretic reasons we will elaborate on in Section~\ref{sec:relations}. In particular, it does not allow for a reliable estimation of the onset of the continuum in the spectrum. In this paper, we will describe a method that allows us to augment the DLCQ approach with trace relations for low-rank gauge groups. This enables us to accurately determine the spectrum of these theories, and to do so at a relatively high numerical resolution, allowing for reliable extrapolation to the continuum limit.
We implement this method explicitly for $N=2, 3$, and $4$. Our results show that, even for such low ranks, the low-lying gluinoball spectra can be well approximated via the expansion in powers of $1/N^2$. In other words, we find that 
\begin{equation}\label{eq:n2_expansion}
M^2= \frac {g^2 N}{\pi} \left(a_0 + a_1 N^{-2} + O(N^{-4}) \right) \,,
\end{equation}
and that for the low-lying gluinoballs $a_1 \ll a_0$. For example, in the  $m_\text{adj} = 0$ theory, for the lightest fermionic bound state $a_0\approx 5.7$ and $a_1\approx 0.0035$;
and for the lightest bosonic bound state $a_0\approx 10.8$ and $a_1\approx 0.013$.
These results are complementary to the earlier studies of the expansions in powers of $1/N^2$, which confirmed its validity for the 
glueballs in 3D and 4D SU(N) gauge theory \cite{Athenodorou:2016ebg,Athenodorou:2021qvs}. Additionally, when $m_\text{adj} = 0$, we find doubly degenerate states starting at the two-body continuum threshold. While one such continuum can be explained by two-particle states, the second cannot, and we interpret this as evidence of screening analogous to the continuum in the single-trace spectrum at large $N$.
 
While our results for $N=3$ and $4$ are qualitatively (and even quantitatively!) quite similar to those in the large $N$ limit, $N=2$ is a special case~\cite{LinThesis}. This is because the $\SU(2)$ gauge theory with an adjoint fermion is equivalent to the $\SO(3)$ gauge theory with a Majorana fermion in the fundamental representation. The bosonic eigenstates of this theory are mesonic, while the fermionic ones are ``baryonic," except the baryon number is valued in $\Z_2$ rather than in $\Z$\@. In the $m_\text{adj}\rightarrow 0$ limit there are no massless mesons or baryons,
unlike in the case of $\SU(N)$ gauge theory with a fundamental Dirac fermion.

The rest of this paper is organized as follows.  In Section~\ref{sec:dlcq}, we outline the DLCQ method for 2D $\SU(N)$ Yang-Mills theory with a Majorana fermion transforming in the adjoint representation. In particular, we show how to find a basis of physical states when $N$ is sufficiently large, and how to compute the action of the mass-squared operator on these states. In Section~\ref{sec:relations}, we show that this basis becomes overcomplete when $N$ is small. We show how to calculate the number of physical states for finite $N$ using representation theory, and how to derive the relations among the overcomplete basis. In Section~\ref{sec:results}, we use this method to compute the spectra for
$\SU(2)$, $\SU(3)$, and $\SU(4)$. We study the dependence on the adjoint mass and find numerical evidence in support of the theory being supersymmetric at finite $N$ when
$m^2_\text{adj}=g^2N/\pi$  \cite{Kutasov:1993gq,popov_supersymmetry_2022}. 
Section~\ref{sec:current_blocks} contains a discussion of the current algebra and exact degeneracies that occur when $m_\text{adj} = 0$.  We end with a discussion of our results in Section~\ref{sec:discussion}.  Various technical details, as well as an alternative method for studying the $\SU(2)$ case, are relegated to the Appendices.

\section{Discretized Lightcone Quantization}\label{sec:dlcq}

Let us start by reviewing the DLCQ setup, following \cite{Bhanot:1993xp,Gross:1997mx,dempsey_exact_2021}.

\subsection{Action and Quantization}

The action of adjoint QCD$_2$ is
\begin{equation}
    S_\text{adj} = \int d^2x\,\Tr\left\lbrack -\frac{1}{4g^2}F_{\mu\nu}F^{\mu\nu} + \frac{i}{2}\overline\Psi \slashed D \Psi - \frac{1}{2}m_\text{adj} \overline\Psi \Psi\right\rbrack.
\end{equation}
Here we use the metric $\eta_{\mu\nu} = \diag\{1, -1\}$, and gamma matrices $\gamma^0 = \sigma_2$, $\gamma^1 = i\sigma_1$ obeying the Clifford algebra $\{\gamma^\mu, \gamma^\nu\} = 2\eta^{\mu\nu}$. The covariant derivative acts as
\begin{equation}
    D_\mu \Psi_{ij} = \partial_\mu \Psi_{ij} + i[A_\mu, \Psi]_{ij} \,,
\end{equation}
where the gauge field $(A_\mu)_{ij}$ is Hermitian and traceless.

We define coordinates $x^\pm  = (x^0 \pm x^1)/\sqrt{2}$, and gauge field components $A_\pm = (A_0 \pm A_1)/\sqrt{2}$. The components of the adjoint fermion are taken to be $\Psi_{ij} = 2^{-1/4}\begin{pmatrix} \psi_{ij} \\ \chi_{ij} \end{pmatrix}$. After fixing the gauge $A_- = 0$, the action becomes
\begin{equation}
    S_\text{adj} = \int d^2x\,\Tr\left(\frac{1}{2g^2}\left(\partial_- A_+\right)^2 + \frac{i}{2}\left(\psi \partial_+ \psi + \chi \partial_- \chi\right) + A_+ J^+ - \frac{i}{\sqrt{2}}m_\text{adj} \chi \psi\right).
\end{equation}
The right-moving component $J^+$ of the $\SU(N)$ current is given by
\begin{equation}
    J^+_{ij} = \psi_{ik}\psi_{kj} - \frac{1}{N}\delta_{ij}\psi_{kl}\psi_{lk} \,.
\end{equation}

If we treat $x^+$ as the time coordinate, then we can integrate out the non-dynamical fields $A_+$ and $\chi$, so that the action becomes
\begin{equation}
    S = \int d^2x\,\Tr\left(\frac{g^2}{2}J^+ \frac{1}{\partial^2} J^+ + \frac{i}{2}\psi \partial_+ \psi + \frac{im_\text{adj}^2}{4}\psi \frac{1}{\partial_-}\psi\right).
\end{equation}
We thus have the lightcone momentum operators
\begin{equation}\label{eq:momenta}
    \begin{split}
        P^+ &= \frac{1}{2}\int dx^-\,\Tr\left(i\psi\partial_- \psi\right),\\
        P^- &= -\int dx^- \,\Tr\left(\frac{g^2}{2}J^+\frac{1}{\partial_-^2}J^+ + \frac{im_\text{adj}^2}{2}\psi\frac{1}{\partial_-}\psi\right) \,.
    \end{split}
\end{equation}

\subsection{Discretization}

To determine the spectrum of the theory, we diagonalize the mass-squared operator $M^2 = 2P^+ P^-$. In DLCQ, this problem is treated numerically by first compactifying the spatial direction $x^-$ on a circle, with the boundary condition
\begin{equation}
    \psi_{ij}(x^-) = -\psi_{ij}(x^- + 2\pi L) \,.
\end{equation}
We then have the mode expansion
\begin{equation}\label{eq:mode_expansion}
    \psi_{ij}(x^-) = \frac{1}{\sqrt{2\pi L}}\sum_{\text{odd }n>0} \left(B_{ij}(n)e^{-inx^-/2L} + B^\dagger_{ij}(n) e^{inx^-/2L}\right) \,,
\end{equation}
where the dimensionless operators $B_{ij}(n)$ and $B^\dagger_{ij}(n)$ obey the algebra
\begin{equation}
    \begin{split}
        \left\lbrace B_{ij}(m), B^\dagger_{kl}(n)\right\rbrace &= \delta_{m,n}\left(\delta_{il}\delta_{kj} - \frac{1}{N}\delta_{ij}\delta_{kl}\right),\\
        \left\lbrace B_{ij}(m), B_{kl}(n)\right\rbrace &= \left\lbrace B^\dagger_{ij}(m), B^\dagger_{kl}(n)\right\rbrace = 0 \,.
    \end{split}
\end{equation}
By substituting \eqref{eq:mode_expansion} into \eqref{eq:momenta}, we can write the lightcone momenta $P^\pm$ in terms of the modes $B_{ij}(n)$ and $B^\dagger_{ij}(n)$. Since $P^+$ and $P^-$ commute, we are free to first fix $P^+ = K/(2L)$ for some integer $K$ and then diagonalize $P^-$ on this sector. One can show that
\begin{equation}
    P^+ = \frac{1}{2L}\sum_{\text{odd }n>0}B^\dagger_{ij}(n)B_{ij}(n) \,,
\end{equation}
so a state with $P^+ = K/(2L)$ is one of the form
\begin{equation}\label{eq:proto_state}
    B_{i_1 j_1}^\dagger(n_1)\cdots B_{i_p j_p}^\dagger(n_p)\ket{0}\qquad\text{where}\qquad \sum_{i=1}^p n_i = K \,.
\end{equation}
To be gauge-invariant, all indices need to be contracted, and so we have a product of traces of $B^\dagger$ operators. For instance, at $K = 7$ the gauge-invariant states are
\begin{equation}\label{eq:k7_basis}
    \begin{aligned}
        \ket{\psi_1} &= \Tr\left(B^\dagger(1)^7\right)\ket{0} &
        \ket{\psi_2} &= \Tr\left(B^\dagger(3) B^\dagger(1) B^\dagger(1) B^\dagger(1) B^\dagger(1)\right)\ket{0} \\
        \ket{\psi_3} &= \Tr\left(B^\dagger(3) B^\dagger(3) B^\dagger(1)\right)\ket{0} &
        \ket{\psi_4} &= \Tr\left(B^\dagger(5)B^\dagger(1) B^\dagger(1)\right)\ket{0} \\
        \ket{\psi_5} &= \Tr\left(B^\dagger(3)B^\dagger(1)\right)\Tr\left(B^\dagger(1)^3\right)\ket{0} \,.
    \end{aligned}
\end{equation}

Similarly, substituting \eqref{eq:mode_expansion} into the expression for $P^-$, we find
\begin{equation}\label{eq:pminus}
\begin{split}
    P^- = &\frac{g^2 L}{\pi}\Bigg\lbrack N\sum_{\text{odd }n>0} \left(\frac{y_\text{adj}}{n} + 4\sum_{\text{odd }m<n} \frac{1}{(m-n)^2}\right)B^\dagger_{ij}(n) B_{ij}(n) \\
    + &2\sum_{\text{odd }n_i > 0} \Bigg\lbrace \delta_{n_1+n_2,n_3+n_4}\Bigg\lbrack \left(\frac{1}{(n_1-n_3)^2} - \frac{1}{(n_1+n_2)^2}\right) B^\dagger_{ik}(n_1) B^\dagger_{kj}(n_2) B_{il}(n_3)B_{lj}(n_4) \\
    &\phantom{\sum_{\text{odd }n_i > 0} \delta_{n_1+n_2,n_3+}\Big\lbrack}+\frac{1}{2}\left(\frac{1}{(n_3-n_2)^2} - \frac{1}{(n_4-n_2)^2}\right)B^\dagger_{ij}(n_1)B^\dagger_{kl}(n_2)B_{il}(n_3)B_{kj}(n_4)\Bigg\rbrack \\
    &+\delta_{n_1,n_2+n_3+n_4}\Bigg\lbrack \left(\frac{1}{(n_3+n_4)^2} - \frac{1}{(n_2+n_3)^2}\right)B^\dagger_{ij}(n_1)B_{ik}(n_2)B_{kl}(n_3)B_{lj}(n_4) + \hc\Bigg\rbrack\Bigg\rbrace\Bigg\rbrack
\end{split}
\end{equation}
where $y_\text{adj} = \frac{\pi m^2_\text{adj}}{g^2 N}$. Note that the terms involving $B^\dagger_{ij}(n_1)B^\dagger_{kl}(n_2)B_{il}(n_3)B_{kj}(n_4)$ were not included in 
\cite{Dalley:1992yy,Bhanot:1993xp,Gross:1997mx,dempsey_exact_2021} because they are ``string-breaking,'' in the sense that when acting on a state of the form \eqref{eq:proto_state} with a single trace, they produce terms involving products of traces. Such terms are suppressed by $1/N^2$ in the large $N$ limit, but they cannot be ignored at finite $N$.

To find the eigenvalues of $M^2$, we might try to compute the matrix elements $M^2_{ij} \equiv \braket{\psi_i|M^2|\psi_j}$ as well as the Gram matrix $G_{ij} = \braket{\psi_i|\psi_j}$, and then solve the generalized eigenvalue problem
\begin{equation}\label{eq:geneig}
    M^2_{ij} v_j = \lambda G_{ij} v_j \,.
\end{equation}
However, computing $G_{ij}$ proves to be computationally infeasible at large $K$, since the inner product between two states each involving $p$ operators involves a sum over $p!$ possible contractions.\footnote{Of course, it is possible that a more clever polynomial-time algorithm exists. For instance, the determinant of an $n\times n$ matrix is na\"ively a sum of $n!$ terms, but can in fact be computed in $\mathcal{O}(n^3)$ time using an LU decomposition, or faster still using Strassen's algorithm or more sophisticated algorithms. The similarity of our problem to the determinant (as opposed to the analogous problem for bosonic operators, which is more similar to the permanent, known to be $\#$P-hard \cite{Valiant:1979}) raises the question of whether a faster algorithm might exist, but we will not answer this question here.}

A simple resolution to this difficulty would be to instead compute the action of $M^2$ in the form
\begin{equation}
    M^2\ket{\psi_i} = \sum_j A_{ji}\ket{\psi_j} \,.
\end{equation}
The coefficients $A_{ji}$ can be computed efficiently. We then have
\begin{equation}
    M^2_{ki} = \sum_j G_{kj} A_{ji} = (GA)_{ki} \,.
\end{equation}
If $G$ were non-singular, then the eigenvalues of \eqref{eq:geneig} would be equal to the eigenvalues of $A = G^{-1} M^2$, and so we could simply diagonalize the matrix $A$.  However, the Gram matrix turns out to be highly singular. For instance, for $K=7$, we have 
\begin{equation}\label{eq:k7_gram}
    G_{K = 7} = \begin{pmatrix} 
        7f_3(N) & 0 & 0 & 0 & 0 \\
        0 & f_2(N) & 0 & 0 & 0 \\
        0 & 0 & f_1(N) & 0 & 0 \\
        0 & 0 & 0 & f_1(N) & 0 \\
        0 & 0 & 0 & 0 & 3f_2(N)
    \end{pmatrix} \,,
\end{equation}
where
\begin{equation}\label{eq:fk}
    f_k(N) = \prod_{j=-k}^k (N-j) \,.
\end{equation}
Since $f_k(N) = 0$ for integer $N \le k$, we see from \eqref{eq:k7_gram} that for $\SU(2)$ there are only two physical states at $K = 7$, and for $\SU(3)$ there are only four. That is, for $N\le 3$, there are null states at $K = 7$ that we should remove before proceeding. As we will show in the following section, the number of these null states increases at higher $K$. Enumerating and removing them is the key technical problem underlying DLCQ for adjoint QCD at finite $N$.

\section{Trace Relations for $\SU(N)$}\label{sec:relations}

The nullity of certain states follows ultimately from the representation theory of $\SU(N)$. For instance, from \eqref{eq:k7_gram} we see that the null states for $\SU(2)$ at $K = 7$ are those with more than three copies of $B^\dagger(1)$ in their expressions in \eqref{eq:k7_basis}. And indeed, since the $B^\dagger(1)$ operators must all be antisymmetrized and they transform in the adjoint representation, which for $\SU(2)$ is three-dimensional, there is no nonzero gauge-singlet combination of more than three of them.

In Section~\ref{sec:counts} we generalize this logic by computing antisymmetric tensor powers of the adjoint representations of $\SU(2)$, $\SU(3)$, and $\SU(4)$. For any fixed $N$, at sufficiently large $K$ the majority of states become null; for $\SU(2)$ this occurs even for very modest $K$. In Section~\ref{sec:ch}, we show how to use the Cayley-Hamilton theorem to derive a complete set of null relations for any fixed $N$. We can use this method to remove null states from the overcomplete bases from Section~\ref{sec:dlcq}.

\subsection{Counting Physical States}\label{sec:counts}

Consider a set of $B^\dagger$ operators in which different momenta appear with multiplicities $p_1,\ldots,p_k$. Then, since identical operators anticommute, the number of gauge-invariant states is given by the number of singlets in
\begin{equation}\label{eq:tensor_power}
    \left(\wedge^{p_1} \adj\right)\otimes \left(\wedge^{p_2} \adj\right) \otimes \cdots \otimes \left(\wedge^{p_k} \adj\right) \,,
\end{equation}
where $\wedge^\circ$ denotes an antisymmetric tensor power.

For example, we can work out what happens in $\SU(2)$ when we have two $B^\dagger(1)$'s and two $B^\dagger(3)$'s. For sufficiently large $N$, there are three independent states:
\begin{equation}\label{eq:1133_basis}
\begin{split}
    \ket{\phi_1} &= \Tr\left(B^\dagger(1) B^\dagger(1) B^\dagger(3) B^\dagger(3)\right)\ket{0},\\
    \ket{\phi_2} &= \Tr\left(B^\dagger(1) B^\dagger(3) B^\dagger(1) B^\dagger(3)\right)\ket{0},\\
    \ket{\phi_3} &= \Tr\left(B^\dagger(1) B^\dagger(3)\right)^2\ket{0} \,.
\end{split}
\end{equation}
However, since $\wedge^2 \adj = \adj = \bf{3}$ in $\SU(2)$, and $\bf{3} \otimes \bf{3} = \bf{1} \oplus \bf{3} \oplus \bf{5}$, we see that there can be only one physical state in $\SU(2)$ formed from these operators. Thus, there must be two null relations among the states in \eqref{eq:1133_basis}.

Indeed, for $\SU(2)$ the computation is always roughly this simple, because $\wedge^1 \adj = \wedge^2 \adj = \textbf{3}$, $\wedge^0 \adj = \wedge^3 \adj = \textbf{1}$, and all higher antisymmetric powers of the adjoint vanish. Thus, the problem reduces to counting the number of singlets in some tensor power of the adjoint. The number of singlets in $\otimes^n \textbf{3}$ in $\SU(2)$ is given by the Riordan number $R_n$ \cite{agarwal_singlets_2020} (sequence \oeis{A005043} in OEIS), defined by the recursion
\begin{equation}\label{eq:riordan_recursion}
    R_n = \frac{n-1}{n+1}\left(2R_{n-1} + 3R_{n-2}\right) \,,
\end{equation}
with $R_0 = 1$ and $R_1 = 0$.

For larger groups we do not have such a simple formula, but we can still use \eqref{eq:tensor_power} explicitly. The antisymmetric powers can be computed using the character recursion formula~\cite{zhou_characters_1989}
\begin{equation}\label{eq:character_recursion}
    \chi\left(\wedge^p R; z_i\right) = \frac{1}{p}\sum_{k = 1}^p (-1)^{k-1} \chi\left(R; z_i^k\right) \chi\left(\wedge^{p-k} R; z_i\right) \,,
\end{equation}
where $\chi(R; z_i)$ is the character of the representation $R$ evaluated at arguments $z_i$. In Table~\ref{tab:antisym} we give all the antisymmetric tensor powers of adjoints in $\SU(3)$ and $\SU(4)$.

\begin{table}[t]
    \centering
    \begin{tabular}{c|c}
        $k$ & $\wedge^k \adj$ in $\SU(3)$ \\
        \hline
        1 & $\rep{8}$ \\
        2 & $\rep 8 \oplus \rep{10} \oplus \overline{\rep{10}}$ \\
        3 & $\rep 1 \oplus \rep 8 \oplus \rep{10} \oplus \overline{\rep{10}} \oplus \rep{27}$ \\
        4 & $2(\rep{8}) \oplus 2(\rep{27})$
    \end{tabular}\vspace{1cm}
    \begin{tabular}{c|>{\centering\arraybackslash}p{0.8\textwidth}}
        $k$ & $\wedge^k \adj$ in $\SU(4)$\\
        \hline
        1 & $\rep{15}$ \\
        2 & $\rep{15} \oplus \rep{45} \oplus \overline{\rep{45}}$ \\
        3 & $\rep 1 \oplus \rep{15} \oplus \rep{20'} \oplus \rep{35} \oplus \overline{\rep{35}} \oplus \rep{45} \oplus \overline{\rep{45}} \oplus \rep{84} \oplus \rep{175}$ \\
        4 & $2(\rep{15})\oplus 2(\rep{20'})\oplus \rep{35} \oplus \overline{\rep{35}} \oplus \rep{45} \oplus \overline{\rep{45}} \oplus 2(\rep{84}) \oplus \rep{105} \oplus 2(\rep{175}) \oplus \rep{256} \oplus \overline{\rep{256}}$ \\
        5 & $\rep{1}\oplus 2(\rep{15}) \oplus \rep{20'} \oplus 3(\rep{45} \oplus \overline{\rep{45}}) \oplus 2(\rep{84})\oplus\rep{105}  \oplus 3(\rep{175}) \oplus 2(\rep{256} \oplus \overline{\rep{256}}) \oplus \rep{280} \oplus \overline{\rep{280}} \oplus \rep{300'}$ \\
        6 & $3(\rep{15}) \oplus \rep{20'} \oplus \rep{35} \oplus \rep{\overline{35}}\oplus 3(\rep{45}\oplus \rep{\overline{45}}) \oplus 3(\rep{84}) \oplus 5(\rep{175}) \oplus 2(\rep{256} \oplus \rep{\overline{256}}) \oplus 2(\rep{280} \oplus \rep{\overline{280}}) \oplus 2(\rep{300'}) \oplus \rep{729}$ \\
        7 & $\rep{1} \oplus 2(\rep{15}) \oplus 3(\rep{20'}) \oplus \rep{35} \oplus \rep{\overline{35}} \oplus 3(\rep{45} \oplus \rep{\overline{45}}) \oplus 4(\rep{84}) \oplus 2(\rep{105}) \oplus 5(\rep{175}) \oplus 3(\rep{256} \oplus \rep{\overline{256}})\oplus \rep{\overline{280}} \oplus \rep{280} \oplus \rep{300'} \oplus 3(\rep{729})$
    \end{tabular}
    \caption{Antisymmmetric powers of the adjoint for $\SU(3)$ and $\SU(4)$. In $\SU(3)$ we have $\wedge^{8-k} \adj = \wedge^k \adj$, and in $\SU(4)$ we have $\wedge^{15-k}\adj = \wedge^k \adj$.}
    \label{tab:antisym}
\end{table}

\begingroup
\setlength{\tabcolsep}{25pt}
\begin{table}
\centering
\begin{tabular}{c|rrrr}
 $K$ & $\SU(2)$ & $\SU(3)$ & $\SU(4)$ & Large $N$ \\
 \hline
 3 & 1 & 1 & 1 & 1 \\
 4 & 1 & 1 & 1 & 1 \\
 5 & 1 & 2 & 2 & 2 \\
 6 & 1 & 2 & 2 & 2 \\
 7 & 2 & 4 & 5 & 5 \\
 8 & 3 & 8 & 9 & 9 \\
 9 & 3 & 10 & 12 & 13 \\
 10 & 3 & 12 & 17 & 18 \\
 11 & 5 & 20 & 30 & 33 \\
 12 & 7 & 31 & 51 & 57 \\
 13 & 7 & 40 & 72 & 85 \\
 14 & 8 & 54 & 108 & 134 \\
 15 & 12 & 80 & 178 & 229 \\
 16 & 15 & 113 & 272 & 375 \\
 17 & 16 & 150 & 395 & 589 \\
 18 & 19 & 200 & 588 & 945 \\
 19 & 25 & 276 & 891 & 1,551 \\
 20 & 31 & 380 & 1,328 & 2,530 \\
 21 & 35 & 502 & 1,927 & 4,057 \\
 22 & 40 & 658 & 2,794 & 6,525 \\
 23 & 51 & 888 & 4,100 & 10,630 \\
 24 & 63 & 1,188 & 5,947 & 17,262 \\
 25 & 70 & 1,544 & 8,476 & 27,799 \\
 26 & 81 & 2,012 & 12,088 & 44,901 \\
 27 & 101 & 2,650 & 17,284 & 72,850 \\
 28 & 120 & 3,463 & 24,506 & 117,981 \\
 29 & 136 & 4,472 & 34,442 & 190,612 \\
 30 & 158 & 5,760 & 48,309 & 308,226 \\
 31 & 190 & 7,448 & 67,690 & 499,167 \\
 32 & 225 & 9,605 & 94,349 & 808,033 \\
 33 & 256 & 12,266 & 130,703 & 1,306,666 \\
 34 & 294 & 15,622 & 180,573 & 2,113,616 \\
 35 & 350 & 19,954 & 249,043 & 3,421,191 \\
 36 & 410 & 25,400 & 342,069 & 5,536,551 \\
\end{tabular}
\caption{The numbers of states for $K = 3$ through 36 for $\SU(2)$, $\SU(3)$, $\SU(4)$, and at large $N$ (in the latter we include all possible multi-trace states). For any fixed value of $N$, we see that for sufficiently large $K$ most of the states at large $N$ become null.}
\label{tab:counts}
\end{table}
\endgroup

To count the number of states at some fixed $K$, we first enumerate all the ways to write $K$ as a sum of odd numbers, and then use the method described above to count the number of singlet states for each of those combinations of $B^\dagger$ operators. The results are given in Table~\ref{tab:counts}.

In Appendix~\ref{app:km}, we show a more powerful method for deriving the results of Table~\ref{tab:counts} that is based on the Kac-Moody algebra discussed in Section~\ref{sec:current_blocks}. Using this method, one can show that the number of physical states $A_{N,K}$ at level $K$ exhibits the Cardy growth
 \es{GotACardy}{
	A_{N,K} \sim \exp\left(\sqrt{\frac{(N^2 - 1)K}{6}}\right) \,.
 }
In contrast, in the large $N$ limit the number of states grows exponentially in $K$ rather than in $\sqrt{K}$. Thus, for any finite $N$, at sufficiently large $K$ almost all the na\"ive gauge-invariant states one could write down are null.

\subsection{Cayley-Hamilton Relations}\label{sec:ch}

We know from the calculations in the previous section that many of the states in DLCQ at large $N$ become null for small $N$\@. However, to calculate the spectrum using DLCQ, we need to know precisely what the null states are.

\subsubsection{Example for $N=2$ and $K=8$}

To illustrate the method we will be employing to determine these null states, let us start with an example. We saw that among the states at large $N$ and at $K = 8$ are those in \eqref{eq:1133_basis}, and that for $N=2$ there are two null linear combinations of these states. As discussed in Section~\ref{sec:dlcq}, one way of determining the null combinations is to explicitly compute the Gram matrix. The Gram matrix for the states in \eqref{eq:1133_basis} is
\begin{equation}
    G_{1133} = \begin{pmatrix}
        N f_1(N) & 0 & 2 f_1(N) \\
        0 & 2(N^2-1)(N^2-4) & 0 \\
        2 f_1(N) & 0 & 2(N^2-1)(N^2-2)
    \end{pmatrix} \,,
\end{equation}
with $f_k(N)$ as in \eqref{eq:fk}. Setting $N = 2$, we see that $\ket{\phi_2} = 0$ and $\ket{\phi_1} - \ket{\phi_3} = 0$.

However, as we already mentioned, computing the Gram matrix is inefficient. We could determine these null relations directly in the following way, without ever needing to calculate an inner product. Let $X_1$ and $X_2$ be elements of $\SU(2)$; then we have
\begin{equation}\label{eq:su2_ch}
    \lbrace X_1, X_2\rbrace = X_1^i X_2^j \lbrace \sigma_i, \sigma_j\rbrace = 2X_1^i X_2^j \delta_{ij} \mathbbm{1} = X_1^i X_2^j \Tr(\sigma_i \sigma_j) \mathbbm{1} = \Tr(X_1 X_2) \mathbbm{1} \,.
\end{equation}

We can find both null states by making careful choices for $X_1$ and $X_2$. For instance, if we take $X_1 = B^\dagger(1)$ and $X_2 = \left\lbrack B^\dagger(1), B^\dagger(3)\right\rbrack - \Tr\left(B^\dagger(1)B^\dagger(3)\right)\mathbbm{1}$, where the latter linear combination was chosen so that $\tr X_2 = 0$, then \eqref{eq:su2_ch} gives
\begin{equation}
    B^\dagger(1)^2 B^\dagger(3) - B^\dagger(3) B^\dagger(1)^2 - 2 B^\dagger(1) \Tr\left(B^\dagger(1) B^\dagger(3)\right) = 0 \,.
\end{equation}
If we then multiply both sides on the right by $B^\dagger(3)$ and take the trace, we find
\begin{equation}
    2\Tr\left(B^\dagger(1) B^\dagger(1) B^\dagger(3) B^\dagger(3)\right) - 2\Tr\left(B^\dagger(1) B^\dagger(3)\right)^2 = 0 \,.
\end{equation}
After acting with this operator on the vacuum, this gives $\ket{\phi_1} - \ket{\phi_3} = 0$.

Likewise, if we take $X_1 = B^\dagger(1)$ and $X_2 = i\left\lbrace B^\dagger(1), B^\dagger(3)\right\rbrace$, we find
\begin{equation}
    B^\dagger(1)^2 B^\dagger(3) + 2 B^\dagger(1) B^\dagger(3) B^\dagger(1) + B^\dagger(3) B^\dagger(1)^2 = 2\Tr\left(B^\dagger(1)^2 B^\dagger(3)\right)\mathbbm{1} \,.
\end{equation}
Multiplying on the right by $B^\dagger(3)$ and taking the trace, the first and third terms on the left cancel and the right hand side vanishes, so we are left with
\begin{equation}
    2 \Tr\left(B^\dagger(1) B^\dagger(3) B^\dagger(1) B^\dagger(3)\right) = 0 \,.
\end{equation}
This implies $\ket{\phi_2} = 0$.

\subsubsection{Null states for $N=2$}

We can in fact derive all the null relations by generalizing the example above. The identity \eqref{eq:su2_ch} is a consequence of the Cayley-Hamilton theorem. Indeed, if we have a general $\SU(2)$ element $A = a_i \sigma_i$, its characteristic equation is
\begin{equation}
    0 = \det(A-\alpha\mathbbm{1}) = \alpha^2 - a_i a_i \,,
\end{equation}
and so the Cayley-Hamilton theorem implies $A^2 = a_i a_i \mathbbm{1}$. Substituting the definition of $A$ and symmetrizing the coefficients gives
 \es{sigmaAnti}{
    \left\lbrace \sigma_i, \sigma_j\right\rbrace = 2\delta_{ij} \mathbbm{1} \,,
 }
and contracting with $X_1^i$ and $X_2^j$ reproduces \eqref{eq:su2_ch}. Similarly, if we contract \eqref{sigmaAnti} with Grassmann numbers, we obtain the identity
 \es{XXOdd}{
    \left\lbrack X_1, X_2\right\rbrack = \Tr\left(X_1 X_2\right)\mathbbm{1}
 }
for any two Grassmann-valued elements of $\SU(2)$. For instance, taking $X_1 = B^\dagger(1)$ and $X_2 = B^\dagger(3)$, and then contracting the resulting identity with $B^\dagger(5)$, we learn that $ \Tr\left(B^\dagger(1) B^\dagger(3) B^\dagger(5)\right) = \Tr\left(B^\dagger(1) B^\dagger(5) B^\dagger(3)\right)$.

While this method is very effective and, as we show below, it generalizes for $N>2$, we in fact do not use it for $N=2$.  For $N=2$ we will use a more efficient method presented in Appendix~\ref{app:su2} that is based on rewriting the $\SU(2)$ gauge theory with an adjoint as an $\SO(3)$ gauge theory with a fundamental.

\subsubsection{Null states for $N>2$}

For $N=2$, the identities \eqref{eq:su2_ch} and \eqref{XXOdd} are simple consequences of well-known properties of the Pauli matrices, but following the same procedure for other groups gives more intricate identities. For instance, for $N=3$ the Cayley-Hamilton theorem implies the following identity for the Gell-Mann matrices \cite{macfarlane_gell-manns_1968}:
 \es{LLL}{
    \lambda_{i}\lambda_j\lambda_{k} + \text{permutations} = \Tr(\lambda_i \lambda_j)\lambda_k + \Tr(\lambda_j \lambda_k)\lambda_i + \Tr(\lambda_k \lambda_i)\lambda_j + \Tr(\lambda_i \{\lambda_j, \lambda_k\})\mathbbm{1} \,.
 }
Contracting this with commuting numbers gives
 \es{RegularXXX}{
    X_1 X_2 X_3 &{}+ \text{permutations} \\
    &= \Tr(X_1 X_2)X_3 + \Tr(X_2 X_3)X_1 + \Tr(X_3 X_1)X_2 + \Tr\left(X_1\{X_2, X_3\}\right) \,,
 }
and with anticommuting numbers gives
 \es{SignedXXX}{
    X_1 X_2 X_3 &{}+ \text{signed permutations} \\
     &= \Tr(X_1 X_2)X_3 + \Tr(X_2 X_3)X_1 + \Tr(X_3 X_1)X_2 + \Tr\left(X_1[X_2, X_3]\right) \,.
 }
Finally, we could have $X_1$ and $X_2$ Grassmann-valued and $X_3$ commuting, giving
 \es{XXXMixed}{
        X_1 X_2 X_3 &{}+ X_1 X_3 X_2 + X_3 X_1 X_2 - X_2 X_1 X_3 - X_2 X_3 X_1 - X_3 X_2 X_1 \\
        &= \Tr(X_1 X_2)X_3 - \Tr(X_2 X_3)X_1 + \Tr(X_3 X_1)X_2 + \Tr\left(X_1\{X_2, X_3\}\right) \,.
 }

For $\SU(4)$, the Cayley-Hamilton identity for the generators $\lambda^{(4)}_i$ is
 \es{LLLL}{
    \lambda_i \lambda_j \lambda_k \lambda_l + \ldots &= -\big(\Tr(\lambda_i \lambda_j)\Tr(\lambda_k \lambda_l) + \ldots\big) + \big(\Tr(\lambda_i \lambda_j) \{\lambda_k, \lambda_l\} + \ldots\big) \\
    &{}+ \big(\Tr(\lambda_i \{\lambda_j, \lambda_k\}) \lambda_l + \ldots\big) + \big(\Tr(\lambda_i\lambda_j\lambda_k\lambda_l) + \ldots\big) \,,
 }
where the dots denote a sum of all inequivalent terms of the same form as the first one. By contracting with commuting or anticommuting numbers, one can derive several identities for $\SU(4)$ elements.

With these basic Cayley-Hamilton identities in hand, we can follow the same procedure as in the example above: consider all possible choices for the $X_i$, and then contract with various operators and multiply by the vacuum to form relations among gauge-invariant states. We can enumerate all possible relations separately for each possible collection of $B^\dagger$ operators. Using the method of Section~\ref{sec:counts} we can determine how many independent null relations there should be, so we can stop searching for new ones when enough have been found.

After finding all the null relations, we can identify a subset of the large-$N$ basis that forms a physical basis for the gauge group in hand. We then use the null relations to write every other basis state in terms of the physical basis. We can compute the action of $P^-$ on the physical basis, which will in general produce other large-$N$ states not in that basis, and then rewrite them in terms of our basis. This gives an effective method for computing the action of $P^-$ on a physical basis for any $N$.

\section{Results}\label{sec:results}

After computing the null relations and using them to calculate the action of $P^-$ on a basis of physical states, we can diagonalize $M^2 = 2P^+ P^-$. In practice we use SLEPc \cite{hernandez_slepc_2005,balay_efficient_1997,balay_petsc_2019,roman_slepc_2021} for this task, although our matrices are small enough that a single core suffices for all the diagonalizations. In Section~\ref{sec:massless} we study the massless point $y_\text{adj} = 0$, at which the theory is in the screening phase \cite{Gross:1995bp,dempsey_exact_2021}. In Section~\ref{sec:massive} we study other values of $y_\text{adj}$, especially $y_\text{adj} = 1$ where the theory is supersymmetric \cite{Kutasov:1993gq,popov_supersymmetry_2022}.

\subsection{Massless Adjoint}\label{sec:massless}

In Figure~\ref{fig:spectra_y0}, we give the spectra of fermions and bosons in the theory with a massless adjoint for $\SU(2)$, $\SU(3)$, $\SU(4)$, and in the large $N$ limit. For $\SU(2)$, these are computed up to $K = 60$ using the alternative method outlined in Appendix~\ref{app:su2}, which permits the calculation of $P^-$ on the physical basis directly without ever invoking the prohibitively large basis of large-$N$ states. For $\SU(3)$ and $\SU(4)$, we use the method described in Section~\ref{sec:ch} to work up to $K = 30$ and $K = 25$ respectively. As an example, working with the $\SU(3)$ theory at $K = 30$ requires finding 302,466 relations, and then diagonalizing $P^-$ on the remaining 5,760 states. Finally, for large $N$ all string-breaking and string-joining terms are suppressed, so it suffices to compute and diagonalize $P^-$ on the single-trace sector and then assemble multi-trace eigenstates from these single-trace building blocks. For this we use the single-trace spectra computed in \cite{dempsey_exact_2021}. The spectrum in Figure~\ref{fig:spectra_y0} differs from those in \cite{dempsey_exact_2021}, because here we also include the multi-trace sectors.

The bosonic spectra of $\SU(3)$ and $\SU(4)$ are similar to that of $\SU(2)$, with a bound state at $M^2 \approx 10.8 \frac{g^2 N}{\pi}$ and a continuum beginning at $M^2 \approx 22.9\frac{g^2 N}{\pi}$. In each case, this continuum is interpreted as the spectrum of two-particle states formed from the fermion ground state at $M^2 \approx 5.7\frac{g^2 N}{\pi}$ \cite{Gross:1997mx,dempsey_exact_2021}. 

The fermion spectrum is also quite similar among $\SU(3)$, $\SU(4)$, and large $N$: in each case we see bound states at $M^2 \approx 5.7\frac{g^2 N}{\pi}$ and $M^2 \approx 17.2\frac{g^2 N}{\pi}$, and a continuum beginning at what appears to be the same mass as the boson continuum. In the $\SU(2)$ theory, however, there are some marked differences. There is no bound state at $M^2 \approx 17.2\frac{g^2 N}{\pi}$; this is because this state at higher $N$ is odd under charge conjugation, but $\SU(2)$ has no complex representations and hence all states are even under charge conjugation. Furthermore, in $\SU(2)$ the continuum does not begin until $M^2 \approx 32.2\frac{g^2 N}{\pi}$. We interpret this as the two-particle continuum formed from the lowest fermion and the lowest boson states, since indeed
\begin{equation}
    \left(\sqrt{5.7}+\sqrt{10.8}\right)^2 \approx 32.2 \,.
\end{equation}
Finally, in $\SU(2)$ there is an additional bound state at $M^2 = 25.4\frac{g^2 N}{\pi}$. 

\begin{figure}
	\centering
	\begin{subfigure}[t]{\textwidth}
		\centering
		{\Large $\SU(2)$}\\
		\includegraphics[width=.75\linewidth]{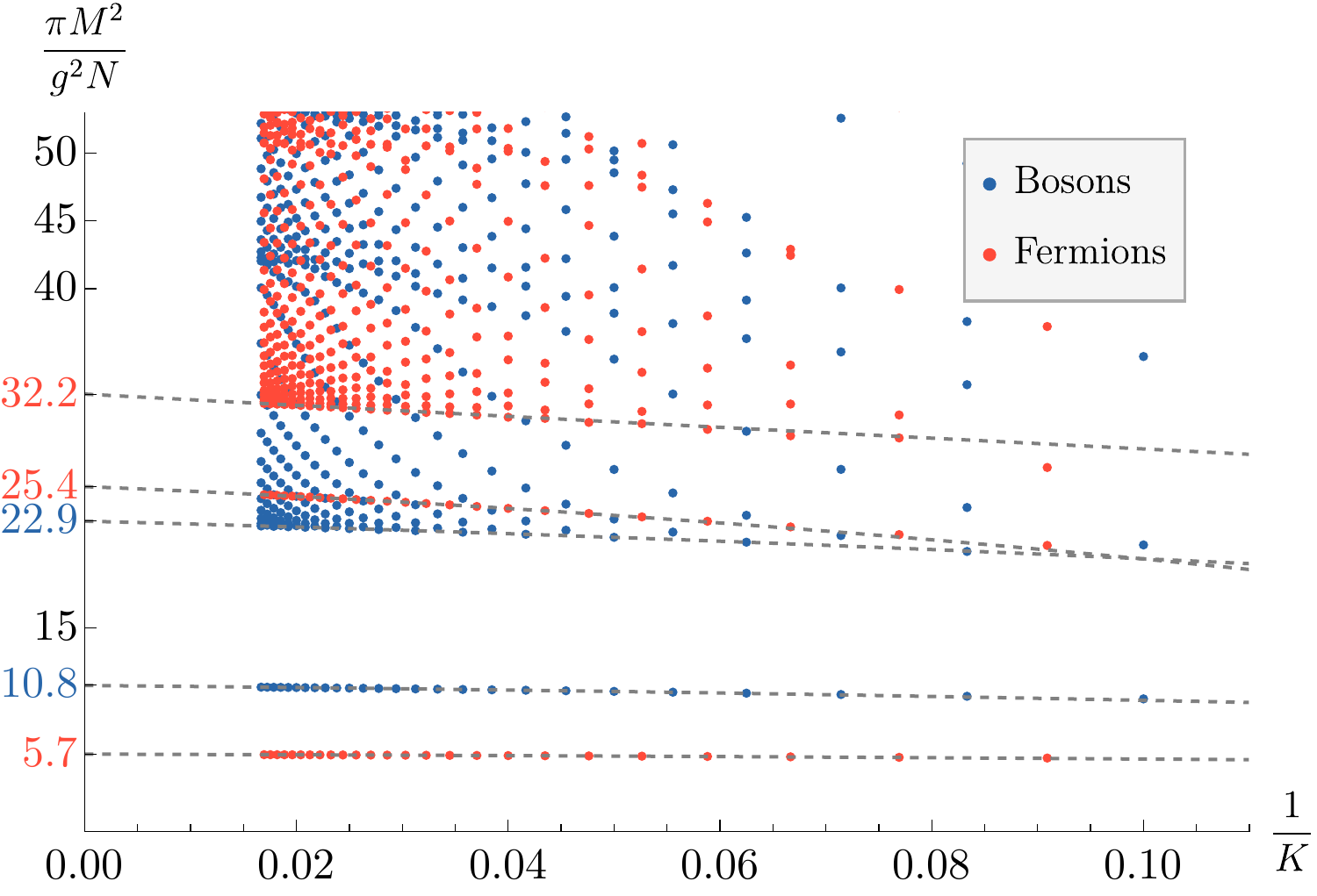}
		\caption{}
		\label{fig:su2_y0}
	\end{subfigure}\\[1em]
	\begin{subfigure}[t]{\textwidth}
		\centering
		{\Large $\SU(3)$}\\
		\includegraphics[width=.75\linewidth]{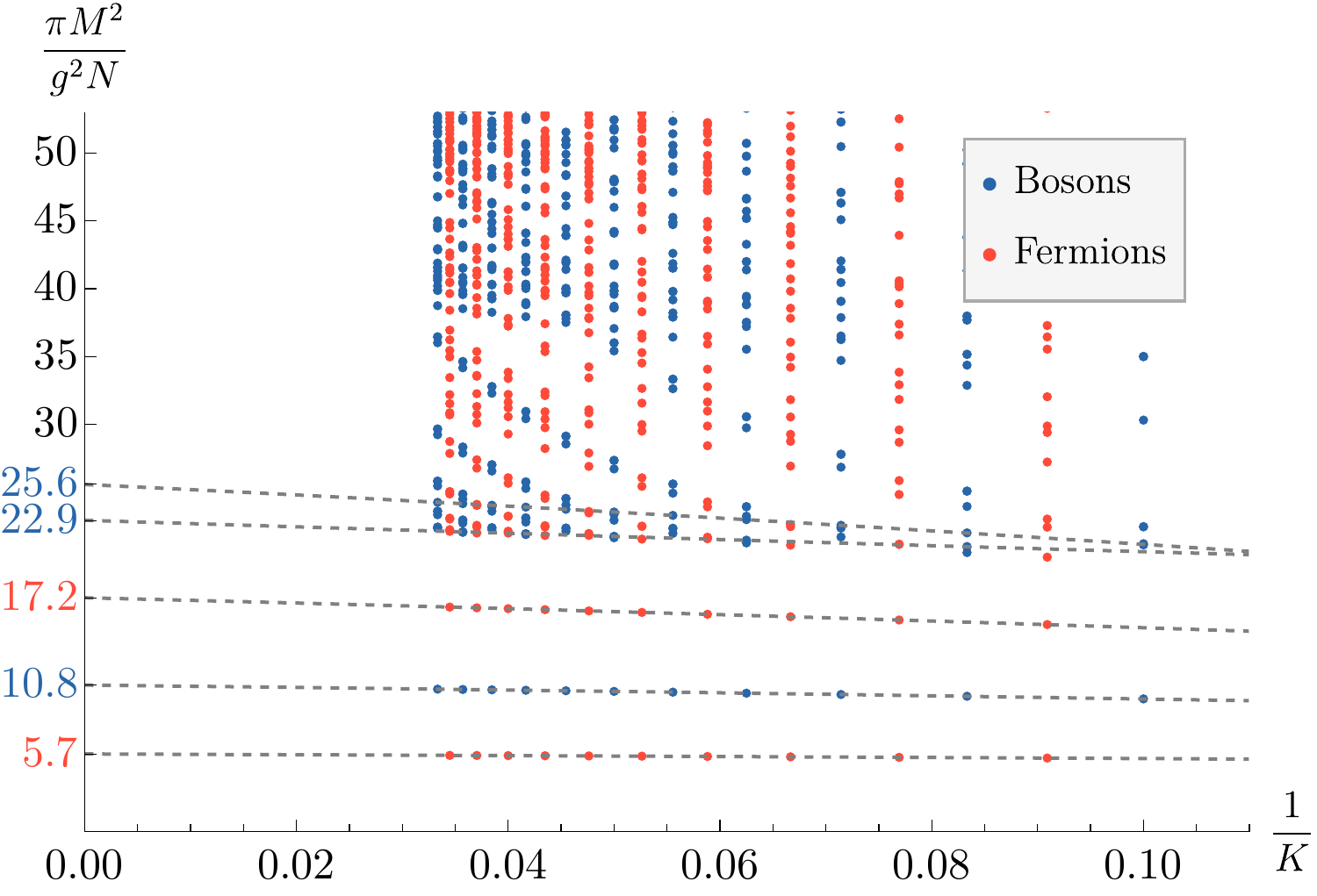}
		\caption{}
		\label{fig:su3_y0}
	\end{subfigure}
\end{figure}
\begin{figure}
	\centering
	\ContinuedFloat	
	\begin{subfigure}[t]{\textwidth}
		\centering
		{\Large $\SU(4)$}\\
		\includegraphics[width=.75\linewidth]{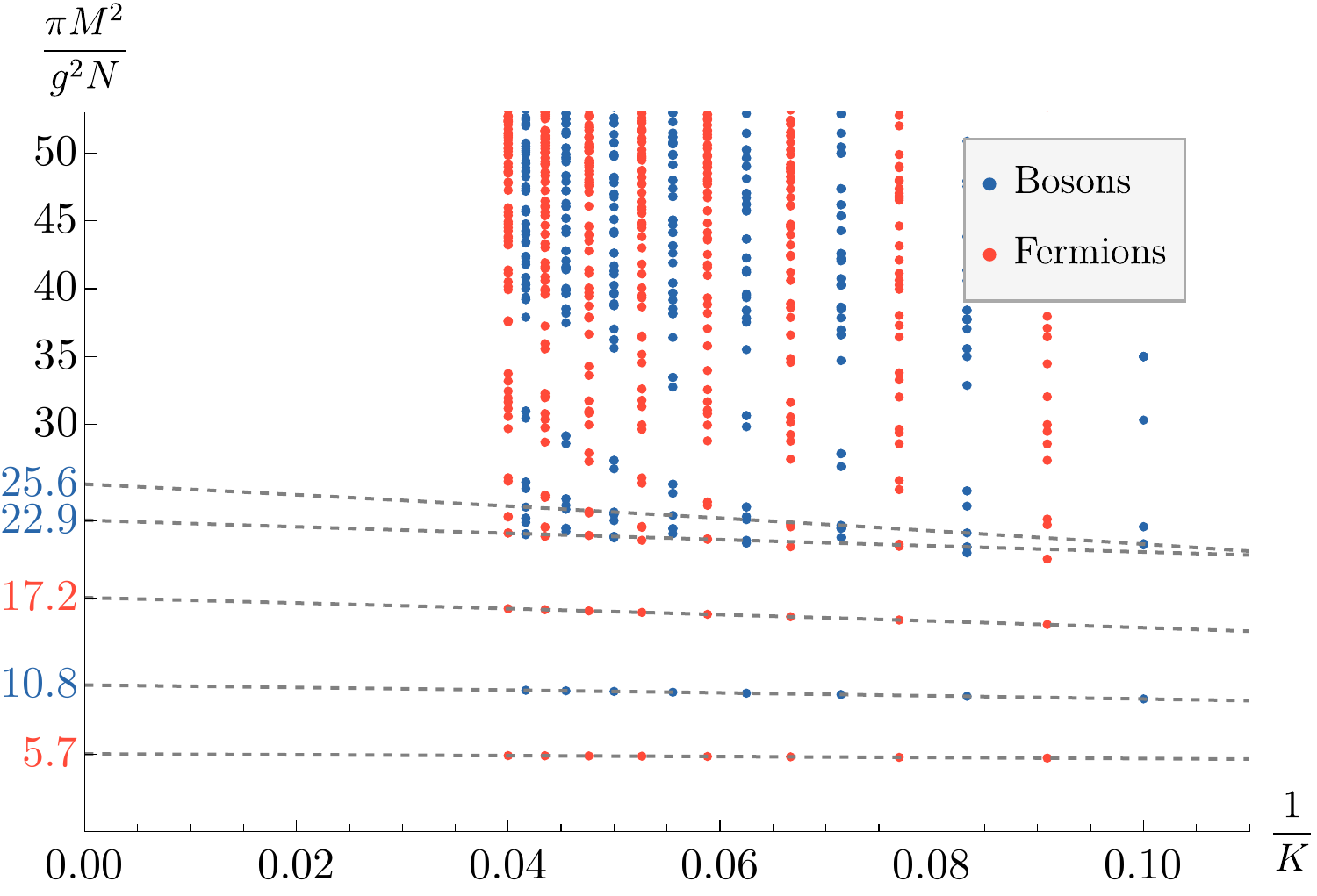}
		\caption{}
		\label{fig:su4_y0}
	\end{subfigure}\\[1em]
	\begin{subfigure}[t]{\textwidth}
		\centering
		{\Large Large $N$}\\
		\includegraphics[width=.75\linewidth]{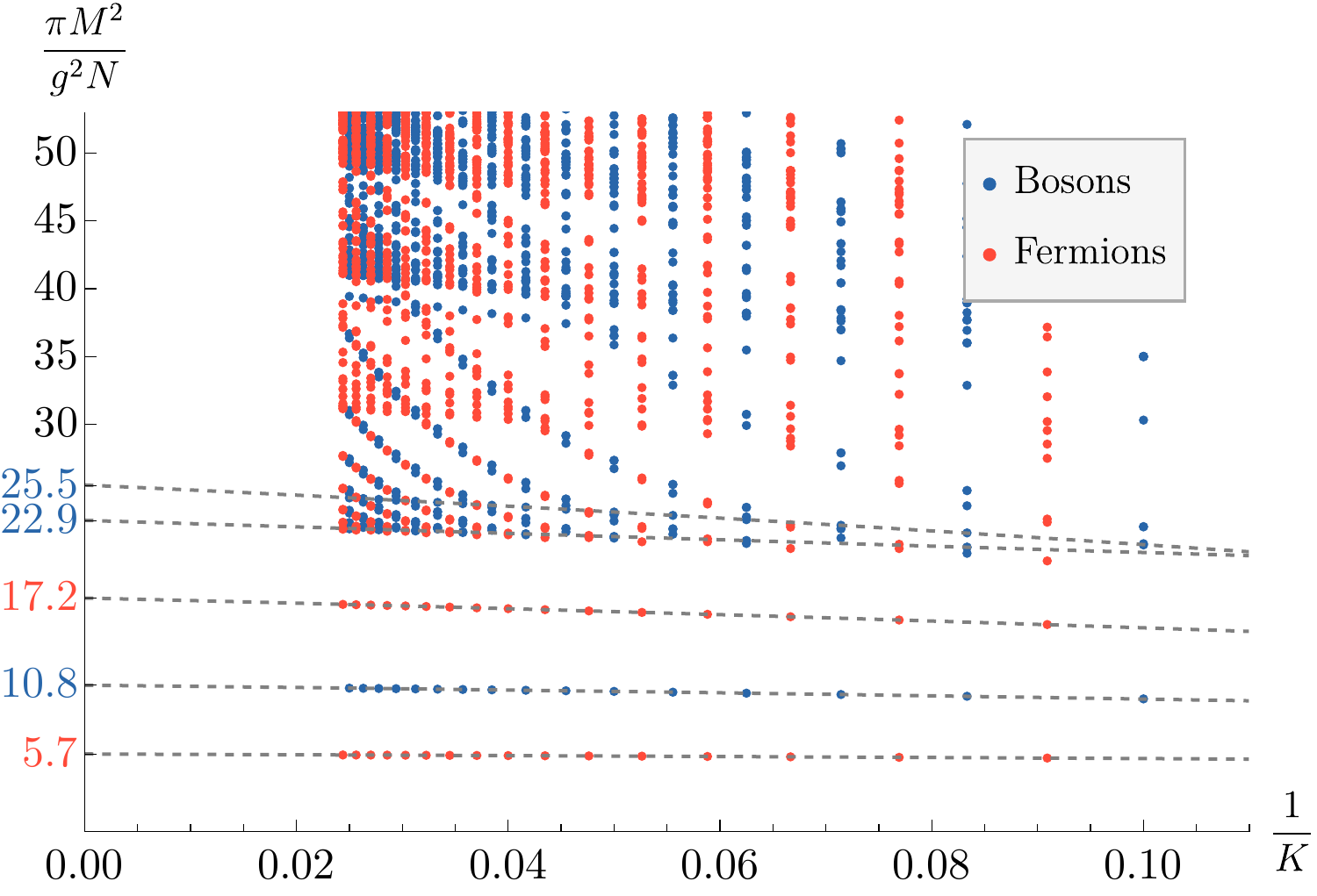}
		\caption{}
		\label{fig:largeN_y0}
	\end{subfigure}%
	
	\caption{The DLCQ spectra at $y_\text{adj} = 0$ for $\SU(2)$, $\SU(3)$, $\SU(4)$. We also exhibit the spectrum in the large $N$ limit where we include the multi-trace states
(the plot is therefore different from \cite{Dalley:1992yy,Bhanot:1993xp,Gross:1997mx,dempsey_exact_2021} where only the single-trace states were exhibited).
The spectrum of bound states below the continuum and the continuum thresholds are similar for $N\ge 3$, with fermion bound states at $M^2 \approx 5.7\frac{g^2 N}{\pi}$ and $M^2 \approx 17.2\frac{g^2 N}{\pi}$ and a boson bound state at $M^2 \approx 10.8\frac{g^2 N}{\pi}$, and continua for both bosons and fermions beginning at $M^2 \approx 22.9\frac{g^2 N}{\pi}$. The $\SU(2)$ theory is distinct, lacking the fermion at $M^2 \approx 17.2\frac{g^2 N}{\pi}$, having an extra fermion at $M^2 \approx 25.5\frac{g^2 N}{\pi}$, and with the fermion continuum threshold at $M^2 \approx 32.2\frac{g^2 N}{\pi}$.}
	\label{fig:spectra_y0}
\end{figure}

\begin{figure}
	\centering
	\includegraphics[width=\linewidth]{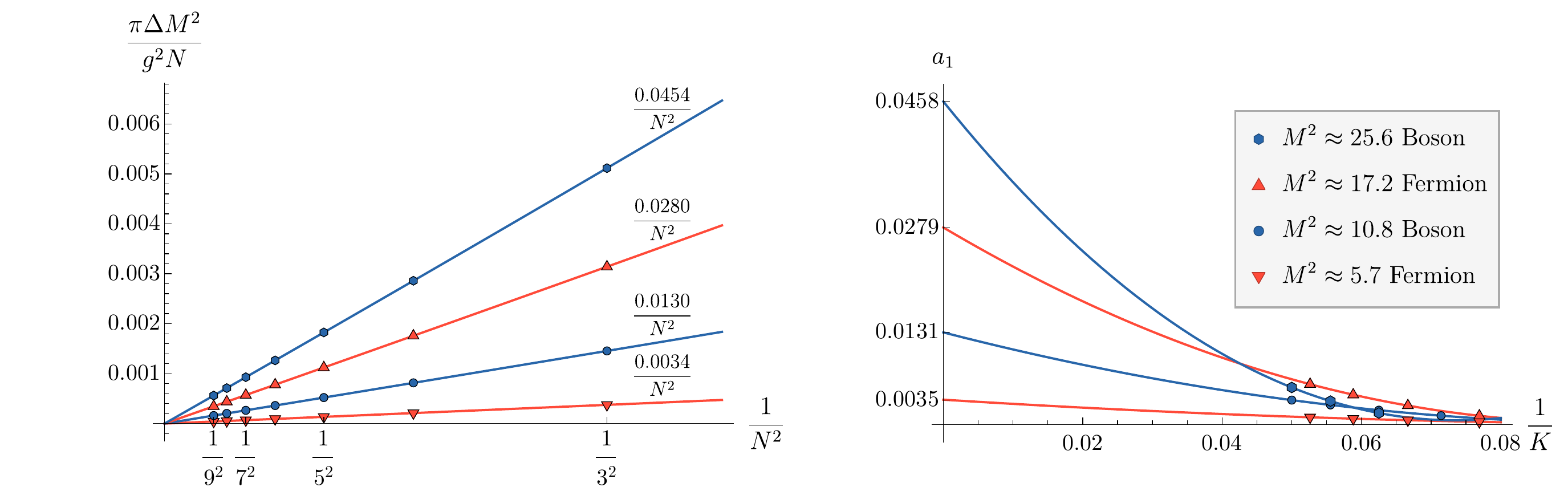}
	\caption{For each of the lowest three bound states in the $N\ge 3$ spectra, we extrapolate the difference between the finite $N$ mass and the large $N$ mass to the continuum, and plot this difference as a function of $N^{-2}$. The leading order correction is seen to be of order $N^{-2}$, with any higher-order corrections much smaller. On the right, we estimate the coefficient $a_1$ of the $N^{-2}$ corrections using eigenvalue perturbation theory at fixed $K$, and then extrapolating the results to large $K$. The results agree well with the fits on the left.}
	\label{fig:n2corrections}
\end{figure}

Given the remarkably close agreement among the low-lying bound state masses for different values of $N$, it is natural to ask what corrections there are to the bound state masses as a function of $N$. Figure~\ref{fig:n2corrections} shows the mass-squared of the lowest three states as a function of $1/N^2$ for several values of $N$\footnote{Here we follow the method of \cite{Antonuccio:1998uz}, ignoring the presence of null states. By comparing to the spectra we obtain after carefully removing null states, we see that this procedure does not change $P^-$ eigenvalues, but only subtracts some of them from the spectrum. Hence, the only error we could be making is if these states themselves are null; but we can be well-assured from Figure~\ref{fig:spectra_y0} that these three bound states are present for $N\ge 3$.}. The corrections are very nearly proportional to $N^{-2}$, with any $N^{-4}$ or higher corrections very small. Furthermore, we see that the magnitudes of the $N^{-2}$ corrections are themselves quite small, explaining why we hardly see any change in these values as a function of $N$ in Figure~\ref{fig:spectra_y0}.

We can also extract the coefficient $a_1$ of the $N^{-2}$ correction (as in \eqref{eq:n2_expansion}) using eigenvalue perturbation theory at a fixed $K$. For any value of $K$, the matrix for $P^-$ can be expressed as
\begin{equation}
	P^-_K = \frac{g^2 N}{\pi}\left(P^-_{0, K} + \frac{1}{N}P^-_{1, K} + \frac{1}{N^2} P^-_{2, K}\right) \,.
\end{equation}
We can similarly expand the eigenvalues $E_i$ of $P^-_K$ in $1/N$:
 \es{EvalExpansion}{
  E_{i, K} = \frac{g^2 N}{\pi} \left(E_{i, 0, K} + \frac{1}{N} E_{i, 1, K} + \frac{1}{N^2} E_{i, 2, K} + \cdots \right)  \,.
 }
At leading order, $E_{i, 0, K}$ are the eigenvalues of $P^-_{0, K}$, but because $P^-_{0, K}$ is not Hermitian, for each eigenvalue we have a left eigenvector $\bra{l_{i, K}}$ and a right eigenvector $\ket{r_{i, K}}$ that can be chosen to be orthonormal $\braket{l_{i, K}|r_{j, K}} = \delta_{ij}$.  The first order correction due to $P^-_{1, K}$ can be computed using a modified expression for first order non-degenerate perturbation theory, and it can be shown to vanish:
 \es{E1}{
  E_{i, 1, K} = \braket{l_{i, K}|P^-_{1, K}|r_{i, K}} = 0 \,.
 }
There are then two sources of $1/N^2$ corrections to the eigenvalues: the first-order corrections from $P^-_{2, K}$, and the second-order corrections from $P^-_{1, K}$:
 \es{E2}{
  E_{i, 2, K} = \braket{l_{i, K}|P^-_{2, K} |r_{i, K}} + \sum_{j\neq i} \frac{\braket{l_{j, K}|P^-_{1, K}|r_{i, K}}\braket{l_{i, K}|P^-_{1, K}|r_{j, K}}}{E_{i, 0, K} - E_{j, 0, K}} \,.
 }
For the $i$th eigenstate, the coefficient $a_1$ defined in \eqref{eq:n2_expansion} is then
 \es{a1FromLimit}{
  a_1 = \lim_{K \to \infty} \left( K E_{i, 2, K} \right)  \,.
 }
We performed this calculation, as well as the large $K$ extrapolation for the four lowest-lying states in the right panel of Figure~\ref{fig:n2corrections}.  The results obtained using this method match very well those obtained using the other order of limits, namely the large $K$ extrapolation followed by the large $N$ extrapolation, which are given in the left panel of Figure~\ref{fig:n2corrections}.

\subsection{Mass Dependence and Supersymmetric Point}\label{sec:massive}

\begin{figure}
	\centering
	\includegraphics[width=.7\linewidth]{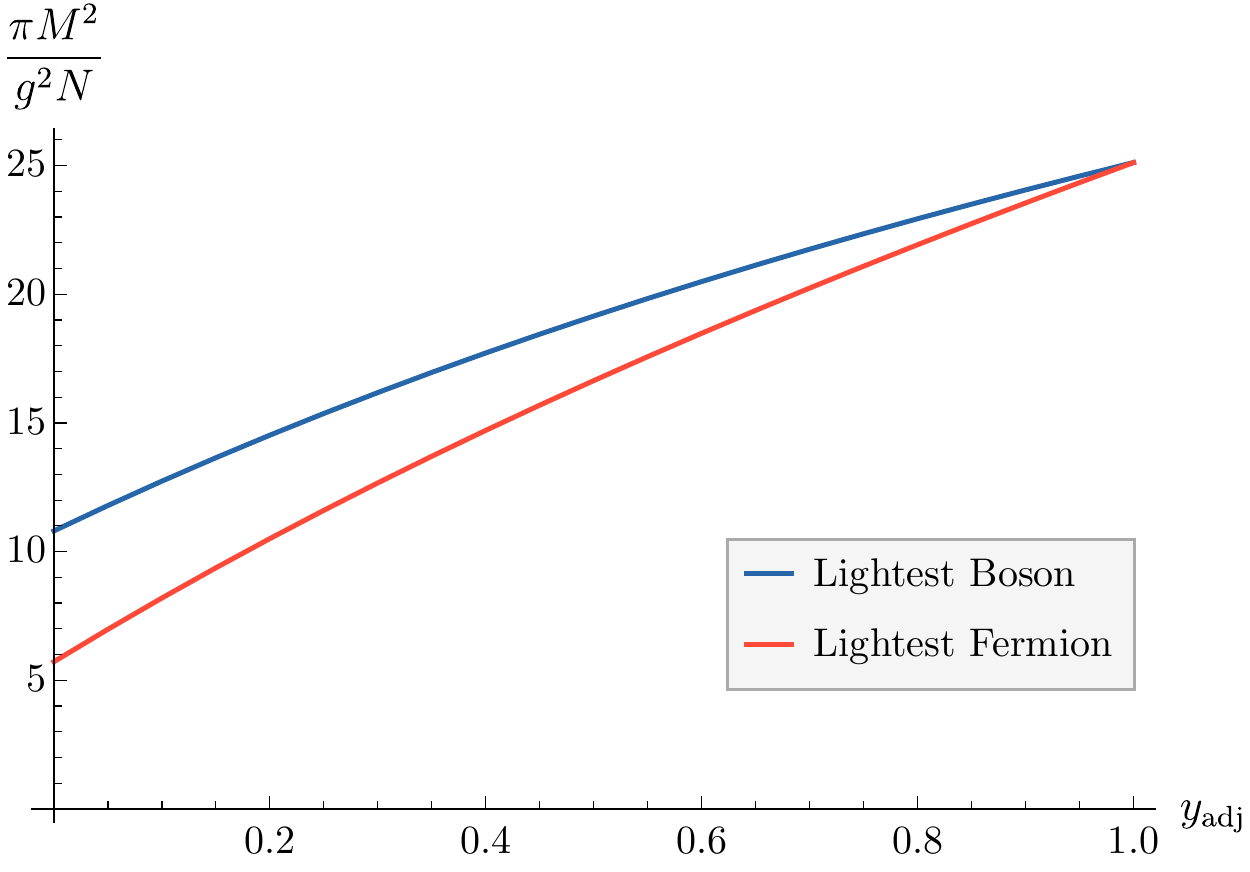}
	\caption{The lightest fermion and boson states in the $\SU(2)$ theory as a function of the adjoint mass parameter $y_\text{adj}$. These states become degenerate at $y_\text{adj} = 1$, a consequence of the emergent supersymmetry at this value.}
	\label{fig:mass_dependence}
\end{figure}

There is no computational difficulty for the mass term in \eqref{eq:pminus}, and so we can freely study these theories at various values of the adjoint mass. In Figure~\ref{fig:mass_dependence}, we show how the masses of the lightest fermion and boson in the $\SU(2)$ theory depend on $y_\text{adj}$. In Table~\ref{tab:mass_dependence}, we give numerical values of the fermion and boson mass gaps for a few specific values of $y_\text{adj}$. The errors are estimated by fitting quadratic functions of $K^{-1}$ to points at various subsets of the values of $K$ where we have computed the spectrum, and looking at the distribution of the extrapolated values we obtain from these fits.

\begin{table}
	\centering
	\begingroup
	\renewcommand{\arraystretch}{1.5}
	\begin{tabular}{|l|cccccc|}
	\hline
 $y_{\text{adj}}$ & 0 & 0.1 & 0.25 & 0.5 & 0.75 & 1 \\
 \hline
 \text{Lowest Fermion} & 5.710(1) & 8.50(5) & 12.15(8) & 17.27(8) & 21.64(7) & 25.57(5) \\
 \text{Lowest Boson} & 10.764(4) & 13.05(5) & 15.95(9) & 19.83(10) & 22.93(8) & 25.58(6) \\
 \hline
	\end{tabular}
	\endgroup
	\caption{Some values of $\frac{\pi M^2}{g^2 N}$ for the lowest fermion and lowest boson in the $\SU(2)$ theory, as given in Figure~\ref{fig:mass_dependence}. Errors are estimated by taking an ensemble of extrapolations to $K\to\infty$ calculated using subsets of 10 or more consecutive points between $K = 30$ and $K = 60$, and taking the standard deviation of the extrapolated values.}
	\label{tab:mass_dependence}
\end{table}

Perhaps the most interesting mass is $m^2_\text{adj} = \frac{g^2 N}{\pi}$, or $y_\text{adj} = 1$. It has long been known that adjoint QCD$_2$ exhibits supersymmetry at this mass \cite{Kutasov:1993gq}, and recently this understanding has been extended to matter in other representations \cite{popov_supersymmetry_2022}. We see this for instance in Figure~\ref{fig:mass_dependence}, with the lightest fermion and boson becoming degenerate at $y_\text{adj} = 1$. In Figure~\ref{fig:spectra_y1}, we give the first numerical demonstration of this supersymmetry in theories with finite $N$, along with the large $N$ spectra at $y_\text{adj} = 1$ assembled from \cite{dempsey_exact_2021}. The fermion and boson masses appear to approach the same values as $K\to\infty$.

\begin{figure}
	\centering
	\begin{subfigure}[t]{\textwidth}
		\centering
		{\Large $\SU(2)$}\\
		\includegraphics[width=.75\linewidth]{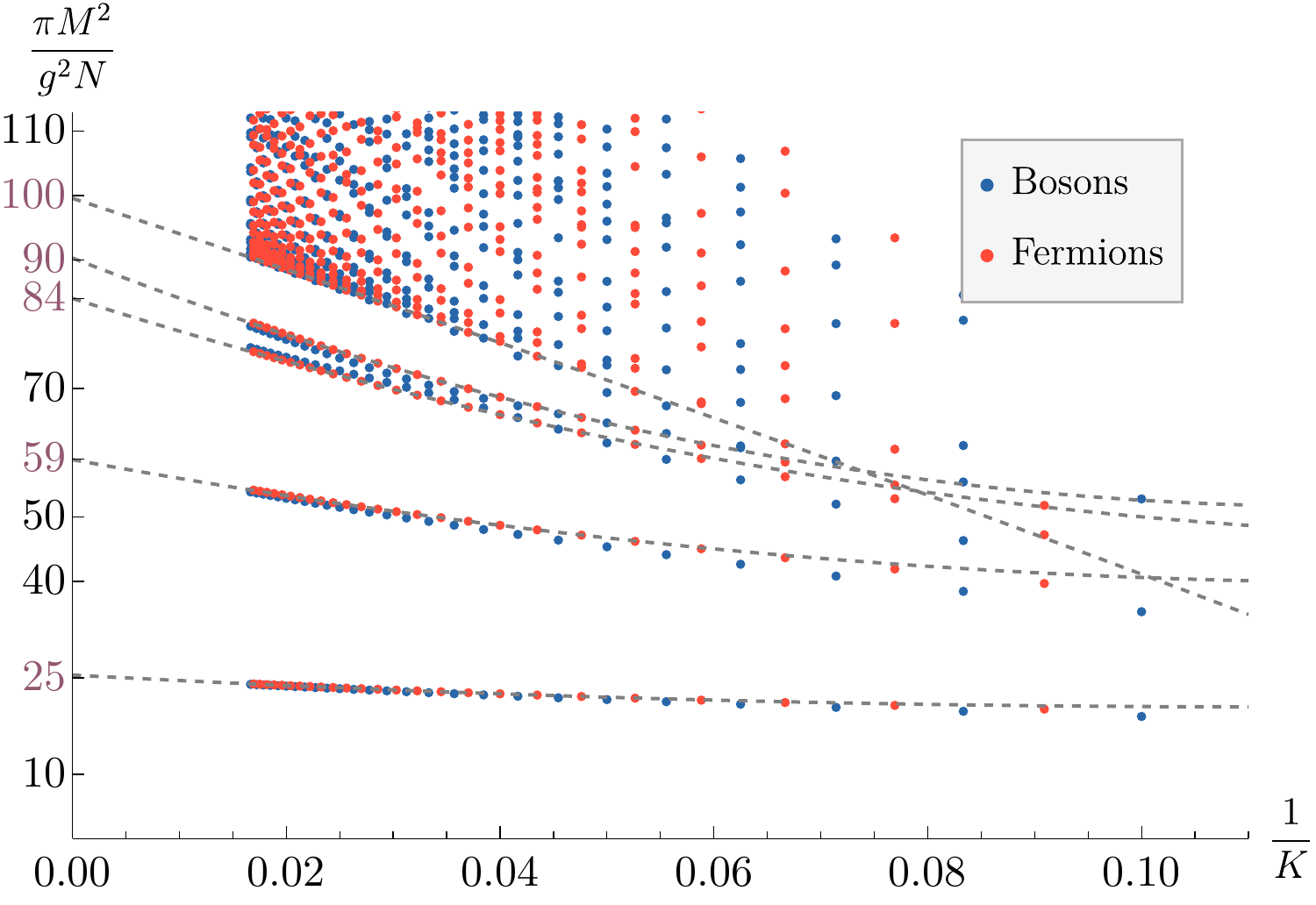}
		\caption{}
		\label{fig:su2_y1}
	\end{subfigure}\\[1em]
	\begin{subfigure}[t]{\textwidth}
		\centering
		{\Large $\SU(3)$}\\
		\includegraphics[width=.75\linewidth]{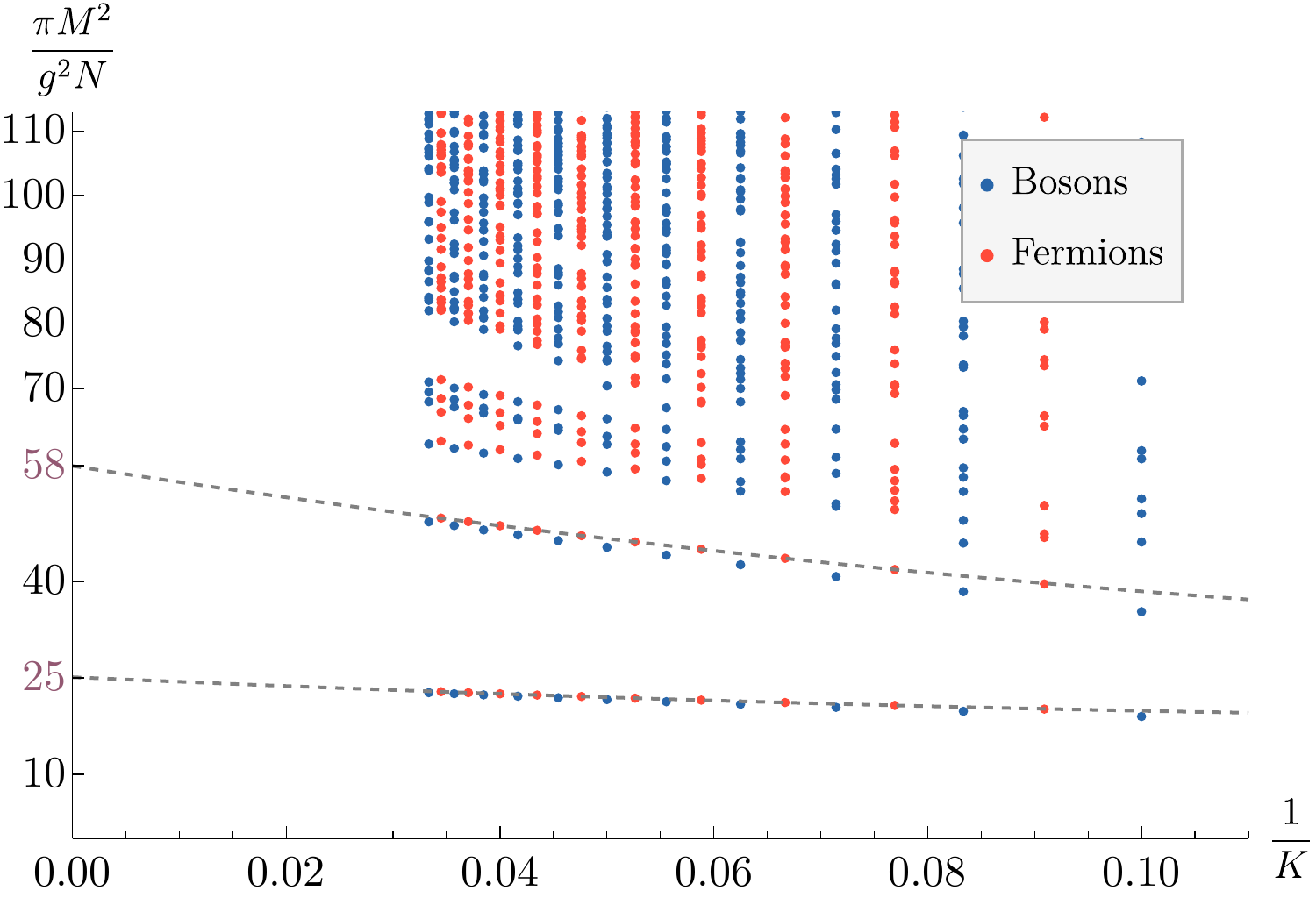}
		\caption{}
		\label{fig:su3_y1}
	\end{subfigure}
\end{figure}
\begin{figure}
	\centering
	\ContinuedFloat
	\begin{subfigure}[t]{\textwidth}
		\centering
		{\Large $\SU(4)$}\\
		\includegraphics[width=.75\linewidth]{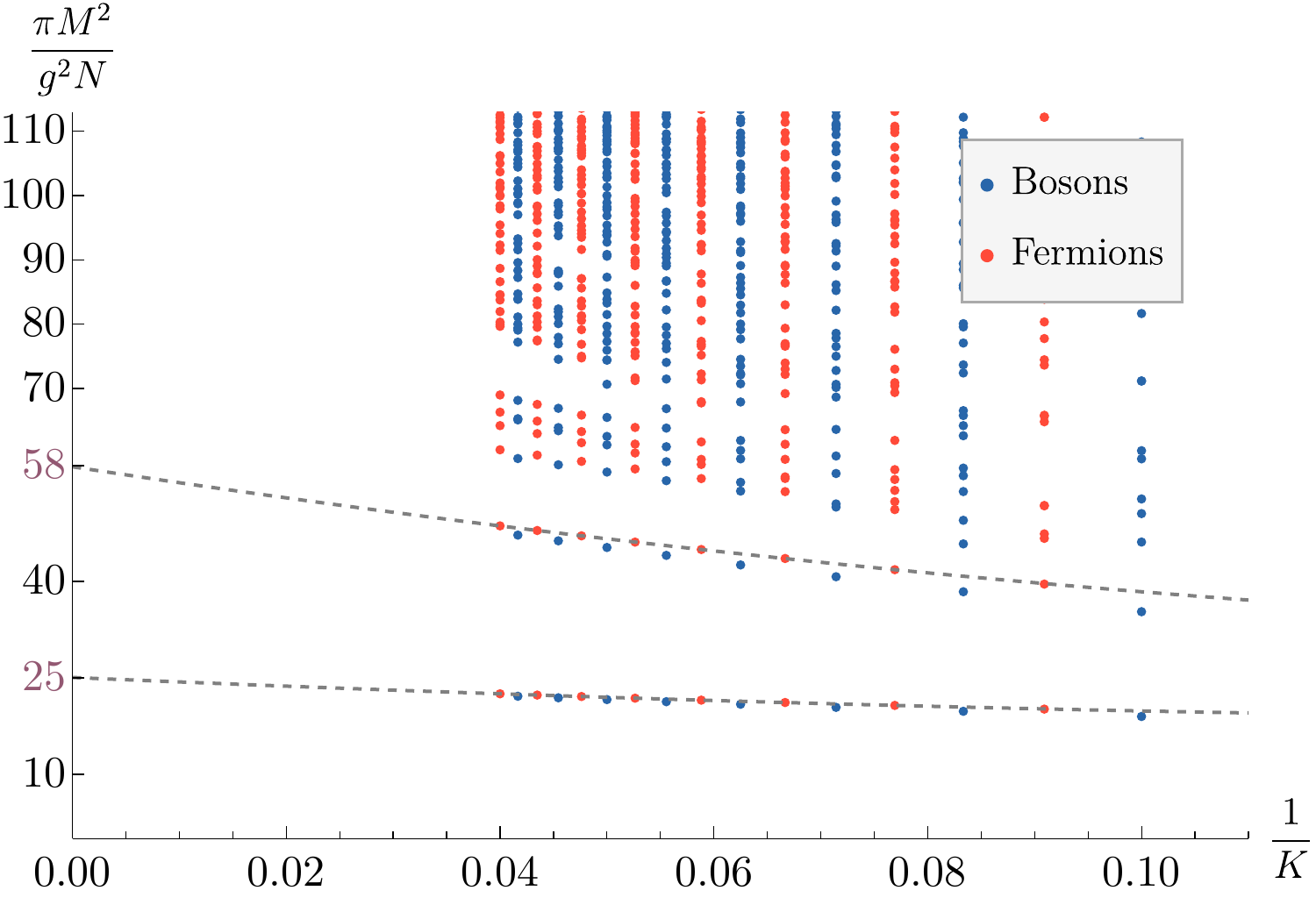}
		\caption{}
		\label{fig:su4_y1}
	\end{subfigure}\\[1em]
	\begin{subfigure}[t]{\textwidth}
		\centering
		{\Large Large $N$}\\
		\includegraphics[width=.75\linewidth]{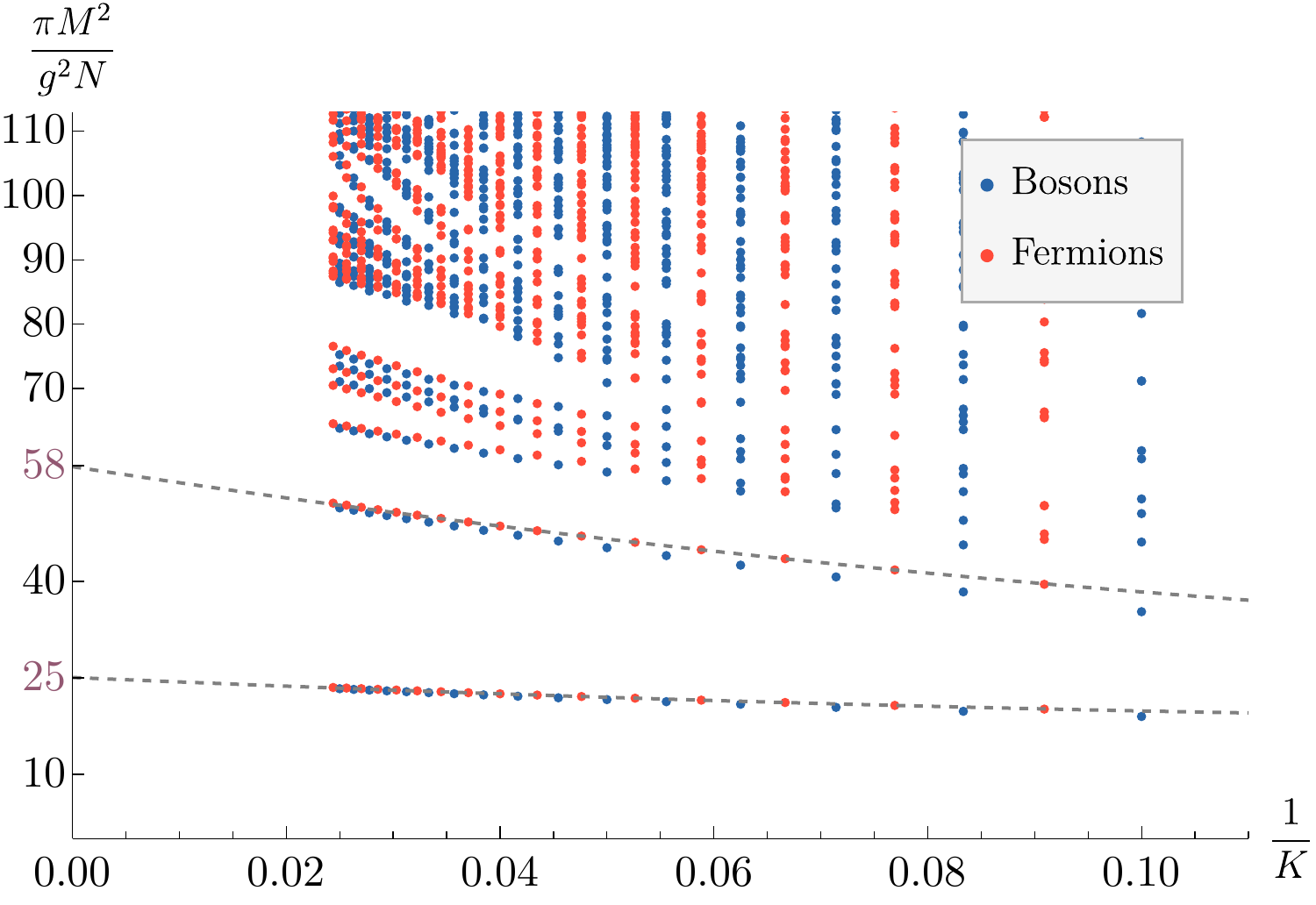}
		\caption{}
		\label{fig:largeN_y1}
	\end{subfigure}%
	\caption{The DLCQ spectra at $y_\text{adj} = 1$ for $\SU(2)$, $\SU(3)$, $\SU(4)$, and large $N$, with maximum $K$ values of 60, 30, 25, and 41, respectively. The spectra exhibit the expected supersymmetry \cite{Kutasov:1993gq,popov_supersymmetry_2022}. The lowest two doublets appear at similar masses for all $N$. For $\SU(2)$, we also extrapolate an additional two bound states and the onset of the two-particle continuum at twice the mass of the lowest doublet.}
	\label{fig:spectra_y1}
\end{figure}

For each value of $N$, we extrapolate the lowest two bound states, which appear at $M^2 \approx 25\frac{g^2 N}{\pi}$ and $M^2 \approx 58\frac{g^2 N}{\pi}$. Since we have especially well-converged spectra for $\SU(2)$, and because the smaller number of states in this theory makes the trajectories easier to distinguish, we also extrapolate two more bound states in this theory at $M^2 \approx 84\frac{g^2 N}{\pi}$ and $M^2\approx 90\frac{g^2 N}{\pi}$. We also see the onset of the continuum at $M^2 \approx 100 \frac{g^2 N}{\pi}$, consistent with the two-particle threshold for the lowest bound state.

\begin{figure}
    \centering
    \includegraphics[width=.9\linewidth]{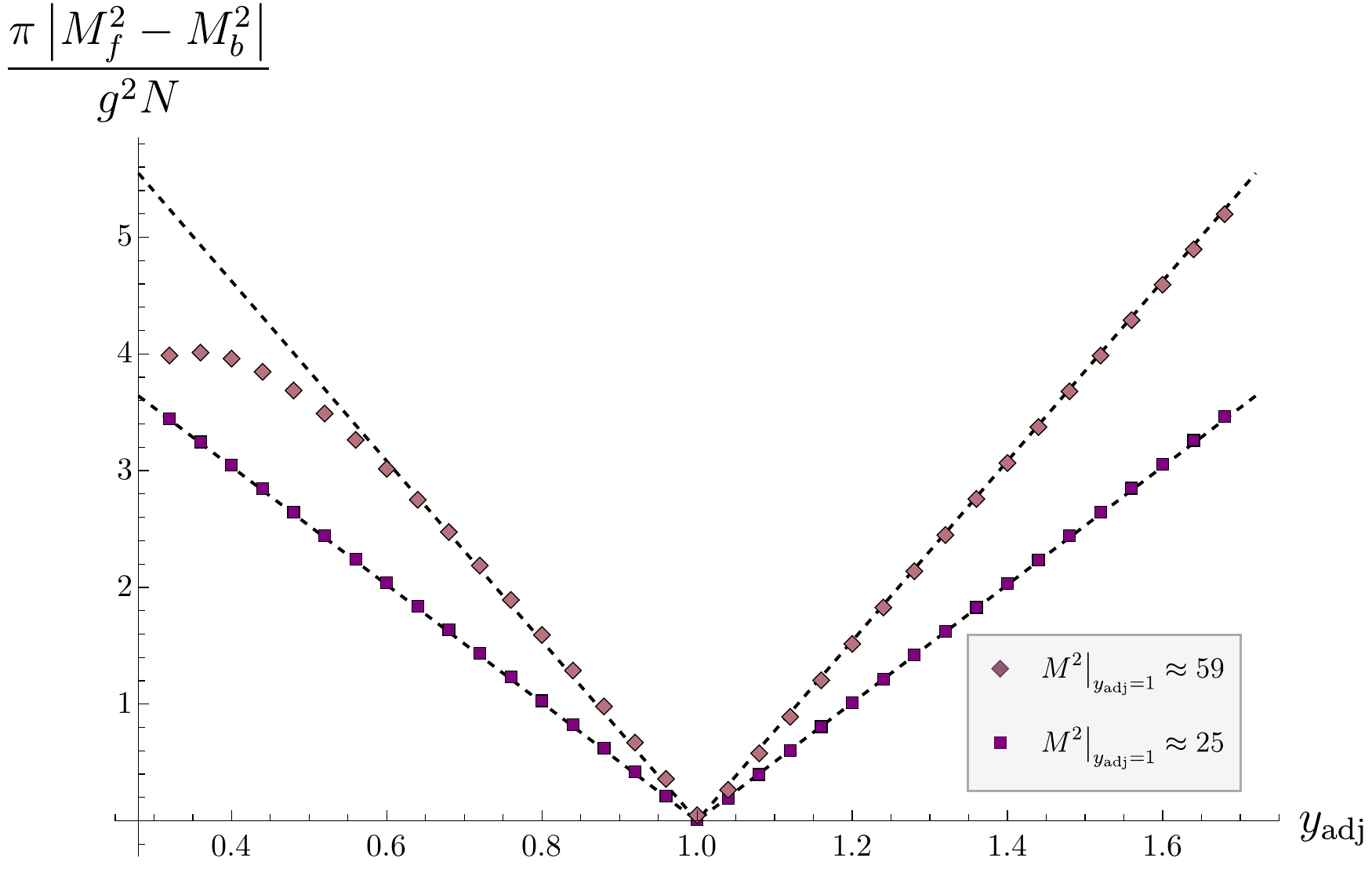}
    \caption{For the lowest two states in the $\SU(2)$ spectrum near the supersymmetric mass $y_\text{adj} = 1$, we plot the absolute difference between the extrapolated continuum boson and fermion masses in the first lowest two doublets as a function of $y_\text{adj}$. We find that they have equal masses (up to extrapolation error) at $y_\text{adj} = 1$. The dashed lines are the prediction of \eqref{eq:splitting}, which shows excellent agreement with the numerics.}
    \label{fig:nearsusy}
\end{figure}

For values of $m_\text{adj}$ near the supersymmetric point, the boson and fermion states that were degenerate at $y_\text{adj} = 1$ should be slightly split. In \cite{boorstein_symmetries_1994}, the splitting is calculated to be
\begin{equation}\label{eq:splitting}
	\frac{\pi}{g^2 N}\left|M^2_f - M^2_b\right| = \sqrt{\frac{\pi}{g^2 N}} M \left|y_\text{adj} - 1\right| + \mathcal{O}\left(\left|y_\text{adj} - 1\right|^3\right) \,,
\end{equation}
where $M^2_f$ and $M^2_b$ are the fermion and boson mass-squareds near $y_\text{adj} = 1$, and $M = \sqrt{\left.M^2_f\right|_{y_\text{adj} = 1}} = \sqrt{\left.M^2_b\right|_{y_\text{adj} = 1}}$ is their common mass at the supersymmetric point. Figure~\ref{fig:nearsusy} shows the excellent agreement between this prediction and our extrapolated continuum $\SU(2)$ spectrum near $y_\text{adj} = 1$.

\section{Current Algebra and Exact Double Degeneracies}\label{sec:current_blocks}

A feature of our results not yet discussed is that, when $m_\text{adj} = 0$, the spectra presented in Figure~\ref{fig:spectra_y0} exhibit some double degeneracies for $N \geq 3$.  These degeneracies were also noticed in \cite{Antonuccio:1998uz}.  As was also mentioned in \cite{dempsey_exact_2021}, the double degeneracies can be understood as a consequence of charge conjugation symmetry and a Kac-Moody algebra structure that makes the light-cone Hamiltonian $P^-$ block diagonal.  Let us now review this explanation and provide more details.

\subsection{Kac-Moody algebra}

When $m_\text{adj} = 0$, $P^-$ can be written solely in terms of the Fourier modes of the current $J^+$   \cite{Kutasov:1994xq,dempsey_exact_2021}.  In terms of the fermionic oscillators, these Fourier modes are
 \es{JModes}{
	J_{ij}(n) = \sum_{m_1 + m_2 = n} \left( B_{ik}(m_1) B_{kj}(m_2) - \frac{1}{N} \delta_{ij} B_{kl}(m_1) B_{lk}(m_2) \right)  \,,
 }
where we used the notation $B_{ij}(-m) = B^\dagger_{ji}(m)$. The modes $J_{ij}(n)$ obey a level $N$ Kac-Moody algebra $\widehat{\mathfrak{su}}(N)_N$
\begin{equation}\label{eq:current_algebra}
	\left\lbrack J_{ij}(n), J_{kl}(m)\right\rbrack = \delta_{kj}J_{il}(n+m) - \delta_{il}J_{kj}(n+m) + N\frac{n \delta_{n+m,0}}{2}\left(\delta_{il}\delta_{kj} - \frac{1}{N}\delta_{ij}\delta_{kl}\right) \,.
\end{equation}
Discretizing the expression for $P^-$ in \eqref{eq:momenta} and using the algebra \eqref{eq:current_algebra} one can write
\begin{equation}\label{eq:pminus_current_positive}
	P^- = \frac{2g^2 N}{\pi} \sum_{\text{even }n > 0} \frac{J_{ij}(-n)J_{ji}(n)}{n^2}\,,
\end{equation}
up to an additive normal-ordering ambiguity that is fixed by requiring that $P^- \ket{0} = 0$.

As explained in \cite{Kutasov:1994xq,dempsey_exact_2021}, the Kac-Moody algebra \eqref{eq:current_algebra} gives a useful way of organizing the states, as well as the spectrum of $P^-$. Indeed, the underlying Hilbert space consisting of all gauge-invariant and gauge non-invariant states organizes itself into highest-weight irreps of the Kac-Moody algebra, also referred to as current blocks.  Each irrep is uniquely specified by the $\SU(N)$ irrep $\rho$ of its Kac-Moody primary state $\ket{\chi}_I$, with $I = 1, \ldots, \dim \rho$.  The Kac-Moody primary is annihilated by all the lowering operators of the current algebra
\begin{equation}\label{eq:primary_def}
	J_{ij}(n)\ket{\chi}_I = 0 \qquad \forall\,  \text{even }n > 0 \,.
\end{equation}
The vacuum is always a Kac-Moody primary, but in general there are others, such as the state $\ket{\chi}_{ij} = B^\dagger_{ij}(1)\ket{0} $, which satisfies \eqref{eq:primary_def}, as one can verify explicitly.  As we will review below, there are precisely $2^{N-1}$ Kac-Moody primaries, each transforming in a different irreducible representation of $\mathfrak{su}(N)$.

In addition to the Kac-Moody primary, each current block contains descendants obtained by acting with the raising operators $J_{ij}(-n)$, with $n>0$, on the primary.  The gauge-invariant states in a current block are those annihilated by $J_{ij}(0)$.

When acting with $P^-$ in the form \eqref{eq:pminus_current_positive} on a descendant of some Kac-Moody primary, we can use the algebra \eqref{eq:current_algebra} to move each lowering operator in $P^-$ all the way to the right where it annihilates the primary.  We then obtain a linear combination of states in the same current block as the one we started with. Hence, $P^-$ is block-diagonal on the current blocks.  

\subsection{List of Kac-Moody blocks}\label{sec:km_list}

To understand how many Kac-Moody blocks there are for a given $N$, note that before gauging we start off with $N^2 - 1$ Majorana fermions, whose Hilbert space can be organized into representations of the $\widehat{\mathfrak{so}}(N^2-1)_1$ Kac-Moody algebra.  There are only two such representations: 
 \begin{enumerate}
  \item  the singlet representation, whose states are bosons and consist of modes of the $\mathfrak{so}(N^2-1)$ currents acting on the vacuum $\ket{0}$; 
  \item the vector representation, whose states are fermions and consist of modes of the $\mathfrak{so}(N^2-1)$ currents acting on the primary $B^\dagger_{ij}(1) \ket{0}$, which transforms in the vector representation of $\mathfrak{so}(N^2 - 1)$.
  \end{enumerate}

We should then decompose these two representations of $\widehat{\mathfrak{so}}(N^2-1)_1$ into representations of the $\widehat{\mathfrak{su}}(N)_N$ algebra \eqref{eq:current_algebra}.  This decomposition was performed in \cite{kac_modular_1988} and used in a related context in \cite{Komargodski:2020mxz,delmastro_infrared_2021}.  The answer is that there are precisely $2^{N-1}$ irreps of $\widehat{\mathfrak{su}}(N)_N$ in this decomposition, each appearing with unit multiplicity.  The Kac-Moody primaries of these irreps transform in $\mathfrak{su}(N)$ irreps $\lambda$ with the following two properties:
 \begin{enumerate}
  \item The Young diagram corresponding to $\lambda$ has at most $N$ columns.  In other words, in Dynkin label notation $\lambda = [\lambda_1 \lambda_2 \ldots \lambda_{N-1}]$, where $\lambda_i$ equals the number of columns of length $i$ in the Young diagram, we have $\sum_{i=1}^{N-1} \lambda_i \le N$.
  \item The highest weight $\lambda$ of the representation is of the form
   \es{lambdaForm}{
    \lambda = w(\rho) - \rho + N \alpha \,,
   }
 for some $w$ and $\alpha$, where $w$ is a Weyl group element of $\mathfrak{su}(N)$, $\rho$ and $w(\rho)$ are the Weyl vector and its image under $w$, and $\alpha$ belongs to the root lattice.  The Weyl group is isomorphic to the permutation group $S_N$, so $w$ is a permutation.  The $\lambda$'s of the form \eqref{lambdaForm} for which $w$ is an even permutation belong to 
 the decomposition of the singlet representation of $\widehat{\mathfrak{so}}(N^2-1)_1$ (i.e.~they give bosons), while those for which $w$ is an odd permutation belong to 
 the decomposition of the vector representation of $\widehat{\mathfrak{so}}(N^2-1)_1$ (i.e.~they give fermions).
 
 \end{enumerate}
As we review in Appendix~\ref{app:km}, one can check this decomposition explicitly from the decompositions of characters of   $\widehat{\mathfrak{so}}(N^2-1)_1$ into characters of $\widehat{\mathfrak{su}}(N)_N$.

Ref.~\cite{Komargodski:2020mxz} determined that the numbers of bosonic and fermionic Kac-Moody blocks are
 \es{NoOfBlocks}{
  \text{\# of bosonic K-M blocks} &= 
   \begin{cases}
    2^{N-2} &\text{if $N$ is even} \,, \\
    2^{N-2} + 2^{(N-3)/2} &\text{if $N$ is odd} \,,
   \end{cases} \\
   \text{\# of fermionic K-M blocks} &= 
   \begin{cases}
    2^{N-2} &\text{if $N$ is even} \,, \\
    2^{N-2} - 2^{(N-3)/2} &\text{if $N$ is odd} \,.
   \end{cases}
 }

Before giving examples, let us note that we can also determine the value $K = K_\lambda$ at which the Kac-Moody primary of representation $\lambda$ occurs.  Due to the fact that $\widehat{\mathfrak{su}}(N)_N  \subset \widehat{\mathfrak{so}}(N^2-1)_1$ is a conformal embedding, the Sugawara stress tensors of $\widehat{\mathfrak{su}}(N)_N$ and $\widehat{\mathfrak{so}}(N^2-1)_1$ agree.  Since the latter algebra has a free field representation in terms of the Majorana fermions, its Sugawara stress tensor is simply $T^{++}$.  Consequently, the Virasoro $L_0$ operator is just $P^+ L$, and therefore $K_\lambda$ is twice the eigenvalue of $L_0$ for the primary of the representation $\lambda$.  This is given by the standard formula \cite{DiFrancesco:1997nk}
\begin{equation}
	K_\lambda = \frac{(\lambda + 2\rho, \lambda)}{2N} \,.
\end{equation}

The explicit decompositions for $N = 2, 3, 4$ along with the corresponding values of $K_\lambda$ are given in Table~\ref{tab:primaries}.

\begin{table}
	\centering
	\begingroup
	\renewcommand{\arraystretch}{1.5}
	\setlength\tabcolsep{.5cm}
	\begin{tabular}{c|ccccc}
		~ & $K_\lambda = 0$ & $K_\lambda = 1$ & $K_\lambda = 2$ & $K_\lambda = 3$ & $K_\lambda = 4$ \\
		\hline
		$\SU(2)$ & $\mathbf{1}$ & $\mathbf{3}$ & ~ & ~ & ~ \\
		$\SU(3)$ & $\mathbf{1}$ & $\mathbf{8}$ & $\mathbf{10}$, $\mathbf{\overline{10}}$ & ~ & ~\\
		$\SU(4)$ & $\mathbf{1}$ & $\mathbf{15}$ & $\mathbf{45}$, $\mathbf{\overline{45}}$ & $\mathbf{35}$, $\mathbf{\overline{35}}$, $\mathbf{175}$ & $\mathbf{105}$
	\end{tabular}
	\endgroup
	\caption{The $\SU(N)$ representations of the Kac-Moody primaries for $N=2, 3, 4$ and the corresponding values $K = K_\lambda$ where they appear.}
	\label{tab:primaries}
\end{table}

\subsection{Degeneracies} \label{sec:Degeneracies}

One can understand the degeneracies in the spectrum as follows.  Since the action of $P^-$ only depends on the level of the Kac-Moody algebra and the structure of each Kac-Moody representation, if a Kac-Moody irrep $\lambda$ were to appear with multiplicity $d_\lambda$, then the corresponding $P^-$ eigenvalues will also have multiplicity $d_\lambda$. Such a situation, however, does not occur for us, because each Kac-Moody irrep appears with unit multiplicity.\footnote{This is, however, a specific property of adjoint QCD$_2$.  For QCD$_2$ with a fermion in a different $\SU(N)$ irrep, the degeneracies of the various Kac-Moody irrep could be greater than one.}

There is another situation which leads to degeneracies.  If there exists a symmetry of $P^-$ that acts as an outer automorphism of the current algebra or of the underlying Lie algebra,\footnote{The outer automorphisms of a classical Lie algebra corresponds to the symmetries of the corresponding Dynkin diagram. The outer automorphisms of the current algebra correspond to the symmetries of the corresponding extended Dynkin diagram that do not leave invariant the extra node.} then the $P^-$ blocks corresponding to the Kac-Moody irreps that are exchanged under the outer automorphism will have degenerate eigenvalues.  For us, the charge conjugation ${\cal C}$ acting as 
 \es{ChargeConj}{
  {\cal C} \psi_{ij} {\cal C}^{-1} = \psi_{ji} \,, \qquad
    {\cal C} J_{ij}^+ {\cal C}^{-1} = J_{ji}^+
 }
is an outer automorphism of $\mathfrak{su}(N)$ of order $2$ (for $N \geq 3$) that commutes with $P^-$, thus leading to doubly degenerate eigenvalues between the blocks corresponding to any complex irrep $\lambda$ and its conjugate $\bar \lambda$.  Of course, complex representations occur only for $N \geq 3$.

From Table~\ref{tab:primaries}, we see that for $\SU(3)$, we expect exact degeneracies among the bosons that are part of the ${\bf 10}$ and $\overline{\bf 10}$ blocks.  For $\SU(4)$, we expect degeneracies among the bosons that are part of the ${\bf 45}$ and $\overline{\bf 45}$, and also degeneracies between the fermions that are part of the ${\bf 35}$ and $\overline{\bf 35}$.

\begin{figure}
	\centering
	\includegraphics[width=\linewidth]{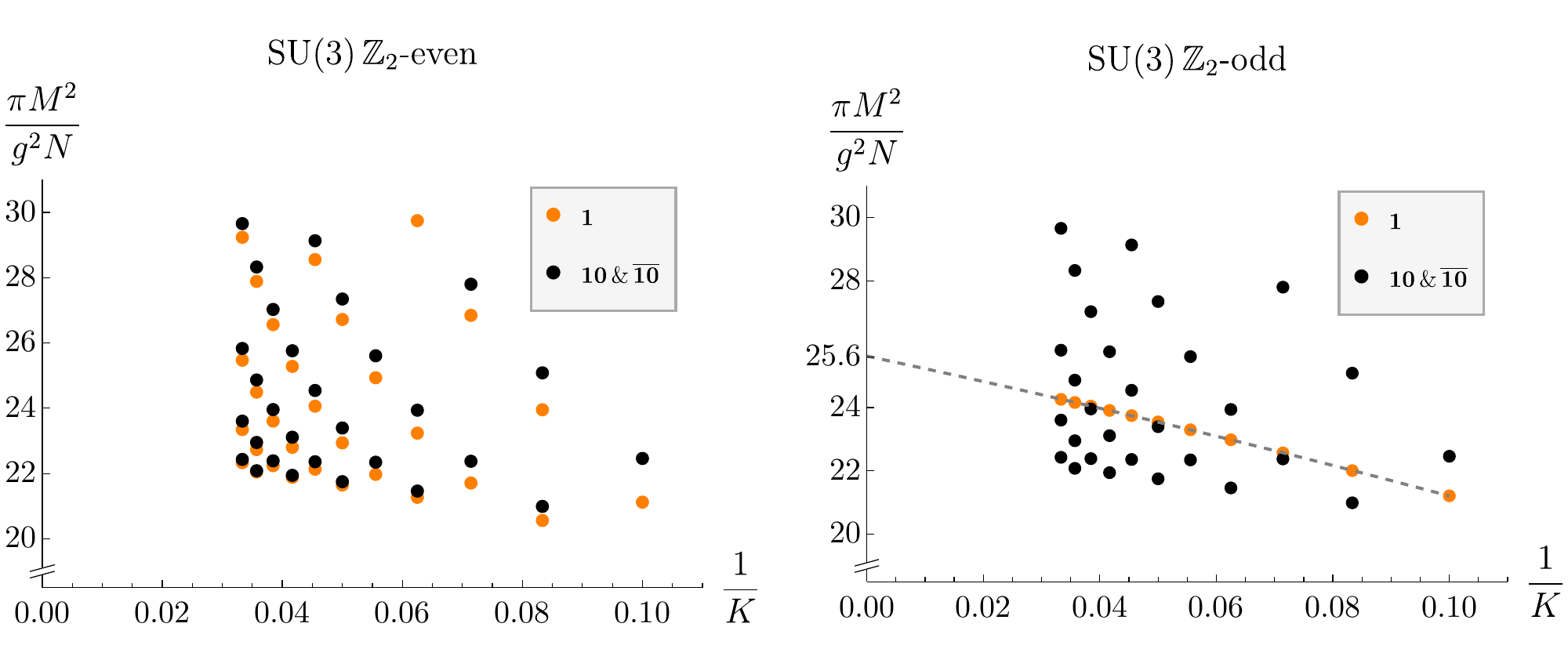}
	\caption{Part of the bosonic $\SU(3)$ spectrum shown in Figure~\ref{fig:spectra_y0}, here split between $\Z_2$-even and $\Z_2$-odd states and with states labeled by the current block they belong to.}
	\label{fig:su3_degeneracy}
\end{figure}

We did not include any information about degeneracies in Figure~\ref{fig:spectra_y0}, but a closer look indeed reveals that some of the states are doubly degenerate while others are non-degenerate.   As an example, let us focus on $\SU(3)$.  Consistent with the discussion above, we do not find any degeneracies in the fermionic spectrum, but we do find degeneracies between $\Z_2$-even and $\Z_2$-odd bosons.  In Figure~\ref{fig:su3_degeneracy}, we plot the bosonic spectrum of the $\SU(3)$ theory for a small range of masses, split according to their charge under the $\mathbb{Z}_2$ charge conjugation symmetry. Some of the masses are repeated in the $\mathbb{Z}_2$-even and $\mathbb{Z}_2$-odd spectra; we identify these as part of the $\mathbf{10}$ and $\mathbf{\overline{10}}$ blocks. All the other eigenvalues are not repeated (they appear either in the $\Z_2$-even or $\Z_2$-odd part of the spectrum, but not both), and hence must belong to the only other bosonic block, the ${\bf 1}$.

After labeling the states by the block to which they belong, a trajectory of vacuum descendants  becomes apparent among the $\mathbb{Z}_2$-odd bosons. This is the state of $M^2 \approx 25.6\frac{g^2 N}{\pi}$ shown in Figure~\ref{fig:spectra_y0}.  We can thus use this method of degeneracies and current blocks to isolate this state from the nearby two-particle continuum, a task that would have been hard to accomplish otherwise.

The exact degeneracies mentioned above can be used to argue for the presence of a continuum of states that survives to finite $N$ and is not explained in terms of a two-body continuum.  Since for all $N \geq 3$, the spectrum in Figure~\ref{fig:spectra_y0} exhibits a massive fermion of squared mass $M_f^2 \approx 5.7\frac{g^2 N}{\pi}$, one expects a two-body continuum to start at $4  M_f^2$ in the bosonic spectrum.  However, as we can see from Figure~\ref{fig:su3_degeneracy} in the $\SU(3)$ case, the trajectories that form this continuum are exactly doubly degenerate even at finite $K$, which implies that the states of the continuum are doubly degenerate too.  This double degeneracy is very surprising, but it follows from charge conjugation symmetry and the Kac-Moody structure.  At large $N$ this degeneracy becomes a degeneracy between single trace and multi-trace states \cite{Kutasov:1994xq,dempsey_exact_2021}.

\section{Discussion}\label{sec:discussion}

In this work, we studied the low-lying spectrum of $\SU(N)$ adjoint QCD$_2$ numerically using DLCQ for $N= 2, 3$, and $4$.  With the adjoint fermion massless, we found a few bound states followed by a continuum.  Surprisingly, we found that the masses of the bound states  receive very small $1/N^2$ corrections.  We also found that the states forming the continuum starting at twice the mass of the lowest fermion exhibit some double degeneracies for $N \geq 3$ even at finite resolution parameter. When the adjoint fermion is massive, there are no double degeneracies but we again find that the $1/N^2$ corrections to the low-lying spectrum are surprisingly small. 
The main challenge we had to overcome was to figure out an efficient way of constructing  a basis of linearly-independent states after taking into account all finite-$N$ trace relations.  As explained in Section~\ref{sec:ch}, we determined the trace relation using a method based on the Cayley-Hamilton theorem.

There are many interesting questions that we leave for the future. As we saw in Section~\ref{sec:current_blocks}, the Kac-Moody algebra provides a good way of classifying the states when $m_\text{adj} = 0$.  Intriguingly, from a representation theoretic perspective, the decomposition into Kac-Moody blocks involves decomposing the trivial and vector representations of $\widehat{\mathfrak{so}}(N^2 - 1)_1$ into representations of $\widehat{\mathfrak{su}}(N)_N$.  The exact same decomposition appears in the related question of determining the vacua of the same theory in equal time quantization \cite{Komargodski:2020mxz}.  It would be very interesting to investigate the relation between these two calculations and determine the precise connection between our mass spectra and the $N$ flux tube sectors of the theory.

While  in Section~\ref{sec:Degeneracies} we performed the decomposition of the states into Kac-Moody blocks explicitly in the $N=3$ case, it would be interesting to perform a similar analysis in the $N=4$ case.  More generally, instead of constructing the states by acting with fermionic oscillators on the Fock vacuum, as we do, one can construct a basis of states directly by acting with the modes of the $\SU(N)$ currents on the Kac-Moody primaries, and compute $P^-$ for each Kac-Moody block separately.  Such an approach was taken at large $N$ in \cite{Trittmann:2000uj,Trittmann:2001dk} for the trivial and adjoint $\widehat{\mathfrak{su}}(N)_N$ blocks.    It would be useful to generalize this approach to finite~$N$.

From a practical point of view, in the $\SU(2)$ case we were able to attain much larger values of $K$ partly because the $\SU(2)$ adjoint theory can also be viewed as an $\SO(3)$ gauge theory with a fundamental Majorana fermion, and in the latter description, we constructed the physical states in a non-redundant way.  This construction involved contracting the fermionic oscillators with the $\SO(3)$ invariant tensors $\delta^{ab}$ and $\epsilon^{abc}$.  It is possible that a similar, more efficient approach for constructing physical states could be generalized to the $\SU(N)$ case, where one can directly contract the fermionic oscillators with the more complicated invariant tensors of $\SU(N)$.  

It would also be interesting to consider various generalizations of the $\SU(N)$ adjoint theories considered here.  One generalization would be to add fermions in the fundamental representation of $\SU(N)$ (quarks). Then one can study the spectrum of baryons and its dependence on the adjoint and fundamental masses. One can also consider the quarks as probes of the adjoint QCD theory by taking their masses to be very large.  From the meson spectrum, one can hopefully extract the quark-antiquark potential, and, following the large $N$ analysis of  \cite{dempsey_exact_2021}, provide further evidence that massless adjoint QCD exhibits screening also at finite $N$.   Another interesting generalization is based on the fact the $\SU(2)$ adjoint theory can also be viewed as an $\SO(3)$ gauge theory with a fundamental Majorana fermion.  This latter theory can be generalized to an $\SO(N)$ gauge theory with a fundamental fermion, and, just like the the $\SO(3)$ case, it exhibits a $\Z_2$-valued baryon number symmetry.

\section*{Acknowledgments}

We thank Ami Katz, Seok Kim, and Fedor Popov for useful discussions.  This work was supported in part by the US National Science Foundation under Grants No.~PHY-2111977 and PHY-2209997, and by the Simons Foundation Grants No.~488653 and 917464. RD was also supported in part by an NSF Graduate Research Fellowship.

\appendix

\section{Alternate Method for $\SU(2)$}\label{app:su2}

The method outlined in Section~\ref{sec:ch} is based on the premise that inner products of states are expensive to compute. However, for $\SU(2)$, there is an alternate approach that allows us to find a physical basis constructively, without needing to compute the null relations in the large-$N$ basis, and moreover to efficiently calculate inner products of these states. This method allows us to compute $P^-$ at higher $K$ for $\SU(2)$ than we could with the method in Section~\ref{sec:ch}. 

Moreover, as shown in Table~\ref{tab:counts}, the size of the physical basis for $\SU(2)$ is vastly smaller than the basis at large $N$. It is thus possible to diagonalize $P^-$ at significantly higher $K$ for $\SU(2)$. Overall, then, the method we describe here enables us to obtain spectra for $\SU(2)$ substantially closer to the continuum limit.  The following discussion is based on~\cite{LinThesis}.

We start from the fact that the $\SU(2)$ adjoint can also be framed as the $\SO(3)$ fundamental. Indeed, we can define
\begin{equation}\label{eq:so3_def}
    B^\dagger_a(n) = \frac{1}{\sqrt{2}}\sigma_a^{ij} B^\dagger_{ji}(n) \,,
\end{equation}
such that
\begin{equation}\label{eq:so3_anticomm}
    \left\lbrace B^\dagger_a(n), B_b(m)\right\rbrace = \delta_{n,m}\delta_{ab} \,.
\end{equation}
As discussed in Section~\ref{sec:counts}, if we have two copies of the same $B^\dagger$ operator then they form an antisymmetric product that also transforms in the $\SO(3)$ fundamental, namely $\epsilon^{abc}B^\dagger_b(n) B^\dagger_c(n)$. Likewise, if we have three copies of the same operator, they are combined in a singlet as $\epsilon^{abc}B^\dagger_a(n) B^\dagger_b(n) B^\dagger_c(n)$. There is no nonzero combination of four or more copies of the same operator, so these are all the cases we need to consider.

Given a set of $B^\dagger$ operators, we can then count how many of them appear with multiplicity one or two. These correspond to free $\SO(3)$ fundamental indices, which must then be contracted with $\SO(3)$ invariant tensors to form a gauge-invariant state. The invariant tensors are $\delta^{ab}$ and $\epsilon^{abc}$, and we can rewrite $\epsilon^{abc}\epsilon^{def}$ as a sum of products of $\delta$-tensors, so we can restrict to at most one $\epsilon$. Using these rules, the number of gauge-invariant states we can write down starting from $m$ $\SO(3)$ fundamentals is
\begin{equation}
    T_m = \begin{cases} 
        \frac{m!}{(m/2)!\times 2^{m/2}} & \text{if $m$ even} \,,  \\
        \frac{m!}{3\times ((m-1)/2)! \times 2^{(m-1)/2}} & \text{if $m$ odd} \,.
    \end{cases}
\end{equation}

In Section~\ref{sec:counts}, we saw that the number of independent singlet states coming from the tensor power of $m$ $\SU(2)$ adjoints is the Riordan number $R_m$. Table~\ref{tab:riordan} compares the number of tensor expressions we can write down, $T_m$, with the Riordan number $R_m$. Up to $m = 10$, which would appear first at $K = 100$, the disparity is not too great.

\begingroup
\setlength{\tabcolsep}{15pt}
\begin{table}[t]
    \centering
    \begin{tabular}{c|cccccccccc}
        $m$ & 1 & 2 & 3 & 4 & 5 & 6 & 7 & 8 & 9 & 10 \\
        \hline
        $R_m$ & 0 & 1 & 1 & 3 & 6 & 15 & 36 & 91 & 232 & 603 \\
        $T_m$ & 0 & 1 & 1 & 3 & 10 & 15 & 105 & 105 & 315 & 945
    \end{tabular}
    \caption{Given $m$ copies of the $\SU(2)$ adjoint, the Riordan number $R_m$ is the number of independent singlets in their tensor product, while $T_m$ is the number of expressions we could construct by contracting with $\SO(3)$ invariant tensors as described in the main text.}
    \label{tab:riordan}
\end{table}
\endgroup

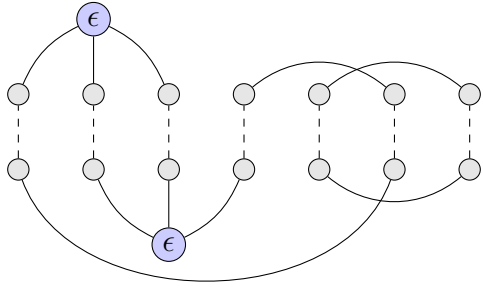
\begin{figure}
    \centering
    \begin{tikzpicture}[mydot/.style={draw,circle,inner sep=.1cm,fill=gray!20},eps/.style={draw,circle,inner sep=.07cm,fill=blue!20}]
\begin{feynman}
\vertex[mydot] (0) at (0, 2) {};
\vertex[mydot] (1) at (1, 2) {};
\vertex[mydot] (2) at (2, 2) {};
\vertex[mydot] (3) at (3, 2) {};
\vertex[mydot] (4) at (4, 2) {};
\vertex[eps] (5) at (1, 3) {$\epsilon$};
\vertex[eps] (7) at (2, 0) {$\epsilon$};
\vertex[mydot] (8) at (0, 1) {};
\vertex[mydot] (9) at (1, 1) {};
\vertex[mydot] (10) at (2, 1) {};
\vertex[mydot] (11) at (3, 1) {};
\vertex[mydot] (12) at (4, 1) {};
\vertex[mydot] (13) at (5, 2) {};
\vertex[mydot] (14) at (6, 2) {};
\vertex[mydot] (15) at (5, 1) {};
\vertex[mydot] (16) at (6, 1) {};
\diagram*{
	(5) --[plain,bend right=20] (0),
	(5) --[plain] (1),
	(5) --[plain,bend left=20] (2),
	(7) --[plain,bend left=20] (9),
	(7) --[plain] (10),
	(7) --[plain,bend right=20] (11),
	(8) --[scalar] (0),
	(9) --[scalar] (1),
	(10) --[scalar] (2),
	(11) --[scalar] (3),
	(12) --[scalar] (4),
	(8) --[plain,bend right=70] (15),
	(3) --[plain,bend left=40] (13),
	(4) --[plain,bend left=40] (14),
	(12) --[plain,bend right=40] (16),
	(15) --[scalar] (13),
	(16) --[scalar] (14)
};
\end{feynman}
\end{tikzpicture}
    \caption{Computing the inner product of two $m = 7$ states. The gray dots represent either a $\SO(3)$ index coming from either one or two $B^\dagger$ operators. Solid lines represent contractions in the individual states, and dotted lines represent $\delta$-tensors coming from \eqref{eq:so3_anticomm}. Calculating the inner product reduces to counting $\SO(3)$ traces and possibly computing a contraction of two $\epsilon$-symbols, which can be accomplished quickly.}
    \label{fig:gluing}
\end{figure}

To identify a physical basis among the $T_m$ tensor contractions for some set of operators, we can in this case just compute the Gram matrix. When calculating the inner product of two such states, we only get nonzero terms when anticommuting two $B^\dagger$'s of the same momentum. For a single $B^\dagger$ we use \eqref{eq:so3_anticomm}, and for two of them, we have
\begin{equation}
    \left\lbrace \epsilon^{acd} B^\dagger_c(n) B^\dagger_d(n), \epsilon^{bef} B_e(m) B_f(m)\right\rbrace = 2\delta_{m,n} \delta^{ab} \,.
\end{equation}
Thus, after anticommuting all $B$'s to the right, we have a single term given by a product of $\epsilon$ and $\delta$ symbols, with at most two $\epsilon$'s since each state can have at most one. We can visually represent the contraction as in Figure~\ref{fig:gluing}. Calculating the contraction amounts to counting loops in a graph, each of which contribute a factor of 3, and possibly computing a contraction of two $\epsilon$-symbols, which gives $\pm 6$.

With the Gram matrix, we can easily identify a subset of $R_m$ of the $T_m$ contractions that are linearly independent, and use those as our basis for the given set of $B^\dagger$ operators. It then remains to compute the action of $P^-$ on that basis. To do so, we should first rewrite $P^-$ in terms of the $\SO(3)$ indices by substituting \eqref{eq:so3_def} into \eqref{eq:pminus}. The result is
\begin{equation}
    P^- = P^-_{1\to 1} + P^-_{2\to 2} + \left(P^-_{1\to 3} + \hc\right) \,,
\end{equation}
where
\begin{equation}
\begin{split}
    \frac{\pi P^-_{1\to1}}{g^2L} &= \sum_{\text{odd }n_i}\left(\frac{2y_\text{adj}}{n}+8\sum_{m\,\mathrm{odd}}^{n-2}\frac{1}{(n-m)^2}\right)B^\dagger_a(n)B_a(n)\,, \\
    \frac{\pi P^-_{2\to2}}{g^2L}&=\sum_{\text{odd }n_i}\Bigg[\left(\frac{3}{2n_{13}^2}-\frac{1}{2n_{14}^2}-\frac{1}{n_+^2}\right)\Bigg(\vcenter{\hbox{\begin{tikzpicture}
\begin{feynman}
\vertex[scale=.8] (0) at (-.5, .5) {\(\scriptsize n_1\)};
\vertex[scale=.8] (1) at (-.5, -.5) {\(\scriptsize n_3\)};
\vertex[scale=.8] (2) at (.5, .5) {\(\scriptsize n_2\)};
\vertex[scale=.8] (3) at (.5, -.5) {\(\scriptsize n_4\)};
\diagram*{
	(0) --[plain,bend right=30, line width=0.15mm] (2),
	(1) --[plain,bend left=30, line width=0.15mm] (3)
};
\end{feynman}
\end{tikzpicture}}}\Bigg)
+\left(\frac{2}{n_{14}^2}-\frac{2}{n_+^2}\right)\Bigg(
\vcenter{\hbox{\begin{tikzpicture}
\begin{feynman}
\vertex[scale=.8] (0) at (-.5, .5) {\(\scriptsize n_1\)};
\vertex[scale=.8] (1) at (-.5, -.5) {\(\scriptsize n_3\)};
\vertex[scale=.8] (2) at (.5, .5) {\(\scriptsize n_2\)};
\vertex[scale=.8] (3) at (.5, -.5) {\(\scriptsize n_4\)};
\diagram*{
	(0) --[plain,bend left=30, line width=0.15mm] (1),
	(2) --[plain,bend right=30, line width=0.15mm] (3)
};
\end{feynman}
\end{tikzpicture}}}
\Bigg)\Bigg]\,, \\
\frac{\pi P^-_{1\to3}}{g^2L}
    &=\sum_{\text{odd }n_i}\left(\frac{1}{n_{34}^2}-\frac{1}{n_{14}^2}\right)\Bigg[\Bigg(\vcenter{\hbox{\begin{tikzpicture}
\begin{feynman}
\vertex[scale=.8] (0) at (-.8, .5) {\(\scriptsize n_1\)};
\vertex[scale=.8] (1) at (.8, .5) {\(\scriptsize n_3\)};
\vertex[scale=.8] (2) at (0, .5) {\(\scriptsize n_2\)};
\vertex[scale=.8] (3) at (0, -.5) {\(\scriptsize n_4\)};
\diagram*{
	(0) --[plain,bend right=50, line width=0.15mm] (2),
	(1) --[plain,bend left=30, line width=0.15mm] (3)
};
\end{feynman}
\end{tikzpicture}}}\Bigg)
-
\Bigg(\vcenter{\hbox{\begin{tikzpicture}
\begin{feynman}
\vertex[scale=.8] (0) at (-.8, .5) {\(\scriptsize n_1\)};
\vertex[scale=.8] (1) at (.8, .5) {\(\scriptsize n_3\)};
\vertex[scale=.8] (2) at (0, .5) {\(\scriptsize n_2\)};
\vertex[scale=.8] (3) at (0, -.5) {\(\scriptsize n_4\)};
\diagram*{
	(0) --[plain,bend right=50, line width=0.15mm] (1),
	(2) --[plain, line width=0.15mm] (3)
};
\end{feynman}
\end{tikzpicture}}}\Bigg)
+
\Bigg(\vcenter{\hbox{\begin{tikzpicture}
\begin{feynman}
\vertex[scale=.8] (0) at (-.8, .5) {\(\scriptsize n_1\)};
\vertex[scale=.8] (1) at (.8, .5) {\(\scriptsize n_3\)};
\vertex[scale=.8] (2) at (0, .5) {\(\scriptsize n_2\)};
\vertex[scale=.8] (3) at (0, -.5) {\(\scriptsize n_4\)};
\diagram*{
	(1) --[plain,bend left=50, line width=0.15mm] (2),
	(0) --[plain,bend right=30, line width=0.15mm] (3)
};
\end{feynman}
\end{tikzpicture}}}\Bigg)\Bigg]\,.
\end{split}
\end{equation}

The line notation indicates a contraction of $B^\dagger$ and $B$ operators in normal order, for instance, 
\begin{equation}
    \left(\vcenter{\hbox{\begin{tikzpicture}
\begin{feynman}
\vertex[scale=.8] (0) at (-.8, .5) {\(\scriptsize n_1\)};
\vertex[scale=.8] (1) at (.8, .5) {\(\scriptsize n_3\)};
\vertex[scale=.8] (2) at (0, .5) {\(\scriptsize n_2\)};
\vertex[scale=.8] (3) at (0, -.5) {\(\scriptsize n_4\)};
\diagram*{
	(1) --[plain,bend left=50, line width=0.15mm] (2),
	(0) --[plain,bend right=30, line width=0.15mm] (3)
};
\end{feynman}
\end{tikzpicture}}}\right) = B^\dagger_a(n_1) B^\dagger_b(n_2) B^\dagger_b(n_3) B_a(n_4) \,.
\end{equation}

We act with $P^-$ on a state simply by writing the state in terms of $B^\dagger$ operators, left-multiplying by $P^-$, and anticommuting all the $B$ operators all the way to the right. At the end of this process, we generically have many terms that are not of the form we have specified, that is, with pairs and triples all combined under an $\epsilon$-symbol. One can derive the following rules that allow us to appropriately recombine operators into the forms we desire:
 \es{Recomb}{
    B^\dagger_a(n) B^\dagger_b(n) B^\dagger_c(n) \delta^{ad} \delta^{be} \delta^{cf} &= \frac{1}{6}\epsilon^{abc}B^\dagger_a(n) B^\dagger_b(n) B^\dagger_c(n) \epsilon^{def} \,, \\
    B^\dagger_a(n) B^\dagger_b(n) B^\dagger_c(n) \epsilon^{abd} \delta^{ce} &= \frac{1}{3}\epsilon^{abc}B^\dagger_a(n) B^\dagger_b(n) B^\dagger_c(n) \delta^{de} \,, \\
    B^\dagger_a(n) B^\dagger_b(n) \delta^{ac} \delta^{bd} &= \frac{1}{2} \epsilon^{eab} B^\dagger_a(n) B^\dagger_b(n) \epsilon^{ecd} \,.
 }

After applying these rules as many times as required, all the operators are combined appropriately, but we may have created additional $\epsilon$-symbols. We could expand the product of $\epsilon$'s into a sum of many products of $\delta$'s, but it is more efficient to simply compute inner products with the several $\epsilon$'s in place. One can show that at most four $\epsilon$ symbols are added when we apply the rules above, so when we compute an inner product, we could have a graph of at most six $\epsilon$'s. There are only a few possibilities, which we can precompute:
\begin{equation}
\begin{split}
    \vcenter{\hbox{\begin{tikzpicture}
\begin{feynman}
\node[dot] (0) at (-.5, 0);
\node[dot] (1) at (.5, 0);
\diagram*{
	(0) --[plain,bend left=30] (1),
	(0) --[plain,bend right=30] (1),
	(0) --[plain] (1)
};
\end{feynman}
\end{tikzpicture}}}=\pm 6 \,, \qquad
    \vcenter{\hbox{\begin{tikzpicture}
\begin{feynman}
\node[dot] (0) at (-.5, .5);
\node[dot] (1) at (.5, .5);
\node[dot] (2) at (-.5, -.5);
\node[dot] (3) at (.5, -.5);
\diagram*{
	(0) --[plain] (1),
	(1) --[plain] (3),
	(3) --[plain] (2),
	(2) --[plain] (0),
	(2) --[plain] (1),
	(0) --[plain] (3)
};
\end{feynman}
\end{tikzpicture}}}&=\pm 6\,,\qquad \vcenter{\hbox{\begin{tikzpicture}
\begin{feynman}
\node[dot] (0) at (-.5, .5);
\node[dot] (1) at (.5, .5);
\node[dot] (2) at (-.5, -.5);
\node[dot] (3) at (.5, -.5);
\diagram*{
	(0) --[plain] (1),
	(3) --[plain] (2),
	(0) --[plain,bend right=30] (3),
	(0) --[plain,bend left=30] (3),
	(2) --[plain,bend left=30] (1),
	(2) --[plain,bend right=30] (1)
};
\end{feynman}
\end{tikzpicture}}}=\pm 12 \,,
\\
    \vcenter{\hbox{\begin{tikzpicture}
\begin{feynman}
\node[dot] (8) at (-.2, .5);
\node[dot] (9) at (-.5, 0);
\node[dot] (10) at (-.2, -.5);
\node[dot] (11) at (0.5, .5);
\node[dot] (12) at (0.5, -.5);
\node[dot] (13) at (.8, 0);
\diagram*{
	(9) --[plain] (10),
	(10) --[plain] (12),
	(12) --[plain] (13),
	(13) --[plain] (11),
	(11) --[plain] (8),
	(8) --[plain] (9),
	(8) --[plain] (12),
	(10) --[plain] (11),
	(9) --[plain] (13)
};
\end{feynman}
\end{tikzpicture}}}=0\,,\quad
\vcenter{\hbox{\begin{tikzpicture}
\begin{feynman}
\node[dot] (8) at (-.2, .5);
\node[dot] (9) at (-.5, 0);
\node[dot] (10) at (-.2, -.5);
\node[dot] (11) at (0.5, .5);
\node[dot] (12) at (0.5, -.5);
\node[dot] (13) at (.8, 0);
\diagram*{
	(9) --[plain] (10),
	(10) --[plain] (12),
	(12) --[plain] (13),
	(13) --[plain] (11),
	(11) --[plain] (8),
	(8) --[plain] (9),
	(8) --[plain] (10),
	(12) --[plain] (11),
	(9) --[plain] (13)
};
\end{feynman}
\end{tikzpicture}}}=\pm 6\,,\quad&
\vcenter{\hbox{\begin{tikzpicture}
\begin{feynman}
\node[dot] (0) at (-.2, .5);
\node[dot] (1) at (.5, .5);
\node[dot] (2) at (.8, 0);
\node[dot] (3) at (.5, -.5);
\node[dot] (4) at (-.2, -.5);
\node[dot] (5) at (-.5, 0);
\diagram*{
	(5) --[plain] (0),
	(0) --[plain,bend right=30] (1),
	(0) --[plain,bend left=30] (1),
	(1) --[plain] (2),
	(2) --[plain] (3),
	(5) --[plain] (2),
	(5) --[plain] (4),
	(4) --[plain,bend right=30] (3),
	(4) --[plain,bend left=30] (3)
};
\end{feynman}
\end{tikzpicture}}}=\pm 24\,,\quad
\vcenter{\hbox{\begin{tikzpicture}
\begin{feynman}
\node[dot] (0) at (-.2, .5);
\node[dot] (1) at (.5, .5);
\node[dot] (2) at (.8, 0);
\node[dot] (3) at (.5, -.5);
\node[dot] (4) at (-.2, -.5);
\node[dot] (5) at (-.5, 0);
\diagram*{
	(5) --[plain,bend right=30] (0),
	(5) --[plain,bend left=30] (0),
	(0) --[plain] (1),
	(1) --[plain,bend right=30] (2),
	(1) --[plain,bend left=30] (2),
	(2) --[plain] (3),
	(5) --[plain] (4),
	(4) --[plain,bend right=30] (3),
	(4) --[plain,bend left=30] (3)
};
\end{feynman}
\end{tikzpicture}}}=\pm24 \,.
\end{split}
\end{equation}
The sign has to be determined in each case by looking in detail at how the indices are contracted.

In summary, for $\SU(2)$ we can efficiently construct a basis of physical states by working in terms of $\SO(3)$ indices. It is then possible to act with $P^-$ directly on this basis, and efficiently compute inner products, allowing us to calculate matrix elements of $P^-$. We can also efficiently calculate the Gram matrix, so that we can find physical mass-squared eigenvalues by solving a problem of the form \eqref{eq:geneig}. This allows us to reach higher values of $K$ for $\SU(2)$ than we could via the method described in Section~\ref{sec:ch}.

\section{Characters and Asymptotics}\label{app:km}

In Section~\ref{sec:current_blocks}, we discussed the Kac-Moody algebra and some of its consequences for the particle spectrum. A key ingredient in this reasoning is the decomposition of the states into Kac-Moody blocks.  In particular, before imposing the gauge-invariance constraint, the states transform in a representation of the $\widehat{\mathfrak{so}}(N^2 - 1)_1$ algebra with two irreducible components (singlet and vector), and these states can then be decomposed under the $\widehat{\mathfrak{su}}(N)_N$ algebra of the gauged $\mathfrak{su}(N)$ currents.  As mentioned in Section~\ref{sec:km_list}, the result of this decomposition can be checked using Kac-Moody characters.  Thus, in Section~\ref{sec:characters} below we give relevant definitions and known formulas for the characters of Kac-Moody algebras.  In Section~\ref{sec:asymptotics} we use these characters to extract the asymptotic behavior of the state counts in Table~\ref{tab:counts}.

\subsection{Characters of Affine Algebras}\label{sec:characters}

The character of a representation $\lambda$ of an affine algebra $\widehat{\mathfrak{g}}_k$ is given by
\begin{equation}\label{eq:character_def}
	\chi^{\widehat{\mathfrak{g}}_k}_\lambda(q, z) = \Tr_\lambda \left(q^{L_0 - \frac{c}{24}} e^{\mathfrak{h}\cdot z}\right)  \,,
\end{equation}
where $z$ is a vector of fugacities of dimension $\rk \mathfrak{g}$, and $\mathfrak{h}$ denotes a basis for the Cartan subalgebra of $\mathfrak{g}$, the underlying Lie algebra of $\widehat{\mathfrak{g}}$. The trace is over all states in the representation $\lambda$. The operator $L_0$ is the zero mode of the Sugawara stress tensor, constructed as
\begin{equation}\label{eq:sugawara_stress}
	T = \frac{1}{2(k + h^\vee_\mathfrak{g})} :j^a j^a: \,,
\end{equation}
where $a = 1,\ldots,\dim \mathfrak{g}$ and $h^\vee_\mathfrak{g}$ is the dual Coxeter number of $\mathfrak{g}$. The central charge of the Virasoro algebra generated by the modes of $T$ is
\begin{equation}
	c = \frac{k \dim \mathfrak{g}}{k + h^\vee_\mathfrak{g}} \,.
\end{equation}
For both $\widehat{\mathfrak{su}}(N)_N$ and $\widehat{\mathfrak{so}}(N^2-1)_1$, we have $c = \frac{N^2-1}{2}$.

The characters can be computed by the Kac-Weyl formula. For $\widehat{\mathfrak{su}}(N)_k$, this takes the form
\begin{equation}\label{eq:kac_weyl}
	\chi^{\widehat{\mathfrak{su}}(N)_k}_\lambda(q, z) = \frac{\sum_{w\in W} \epsilon(w) \Theta_{k+N}(A_{N-1}, w(\lambda + \rho)\mid q, z)}{\sum_{w\in W} \epsilon(w) \Theta_{N}(A_{N-1}, w(\rho)\mid q, z)}\,,
\end{equation}
where $W$ is the Weyl group of $\mathfrak{su}(N)$ (which is the permutation group $S_N$), $\epsilon(w)$ is the signature of an element of this group, and $A_{N-1}$ is the coroot lattice of $\mathfrak{su}(N)$. The theta functions are defined by
\begin{equation}
	\Theta_\kappa(\Lambda, v\mid q, z) = \sum_{x\in \Lambda} q^{\frac{1}{2}\kappa \left(x + \kappa^{-1} v\right)^2} e^{\kappa\left(x + \kappa^{-1} v\right)\cdot z} \,,
\end{equation}
where $\Lambda$ is any lattice and $v$ is a shift vector in the same space as $\Lambda$. At level $k$, the only weights $\lambda = [\lambda_1, \ldots, \lambda_{N-1}]$ that can be the highest weight of a unitary representation are those for which $\sum_{i=1}^{N - 1} \lambda_i \le k$.

For example, in $\widehat{\mathfrak{su}}(2)_2$, there are three unitary representations with $\lambda = 0, \omega, 2\omega$, where $\omega$ is the fundamental weight. Since all our states are built from adjoint fermions, we can only have representations of $N$-ality 0, so we can focus on $\lambda = 0$, the singlet, and $\lambda = 2\omega$, the adjoint. The characters of these representations according to \eqref{eq:kac_weyl} is
\begin{equation}\label{eq:su2_characters}
\begin{split}
	q^{1/16}\chi^{\widehat{\mathfrak{su}}(2)_2}_{\mathbf{1}} &= \mathbf{1} + \mathbf{3} q + \left(\mathbf{1} + \mathbf{3} + \mathbf{5}\right)q^2 + \left(\mathbf{1} + 3(\mathbf{3}) + \mathbf{5}\right)q^3 + \left(3(\mathbf{1}) + 4(\mathbf{3}) + 3(\mathbf{5})\right)q^4 + \ldots\,, \\
	q^{1/16}\chi^{\widehat{\mathfrak{su}}(2)_2}_{\mathbf{3}} &= \mathbf{3} q^{1/2} + \left(\mathbf{1} + \mathbf{3}\right)q^{3/2} + \left(\mathbf{1} + 2(\mathbf{3}) + \mathbf{5}\right)q^{5/2} + \left(2(\mathbf{1}) + 3(\mathbf{3}) + 2(\mathbf{5})\right) q^{7/2} + \ldots\,,
\end{split}
\end{equation}
where we used the shorthand notation ${\bf r}$ to denote the $\mathfrak{su}(2)$ representation of dimension $r$.

A formula very similar to \eqref{eq:kac_weyl} holds for other affine algebras, but for $\widehat{\mathfrak{so}}(N^2 - 1)_1$ we will not need such a formula. Indeed, we construct the $\widehat{\mathfrak{so}}(N^2-1)_1$ algebra by forming currents from the adjoint fermion components $\psi^i$, with $i = 1, \ldots, N^2 - 1$:
\begin{equation}
	j^a = \frac{1}{2}\psi^i T^a_{ij} \psi^j \,,
\end{equation}
with $T^a_{ij}$ the fundamental representation matrices of $\mathfrak{so}(N^2-1)$. We then see that
\begin{equation}
	\Braket{0|j^a(2) j^b(-2)|0} = \frac{1}{4}T^a_{ij} T^b_{kl}\left(\delta^{il}\delta^{kj} - \delta^{ik}\delta^{jl}\right) = \frac{1}{2}\Tr\left(T^a T^b\right) = \delta^{ab} \,,
\end{equation}
which shows that the algebra is at level 1. The stress tensor \eqref{eq:sugawara_stress} is normalized so that the currents have dimension 1, meaning
\begin{equation}
	\left\lbrack L_0, j^a(n)\right\rbrack = -\frac{n}{2} j^a(n) \,.
\end{equation}
Comparing this with $P^+$, we see that $L_0 = P^+ L$, where $L$ is the circle length. This is the quantity we denote by $\frac{K}{2}$. Hence, the characters of $\widehat{\mathfrak{so}}(N^2-1)_1$ are simply counting the states appearing at each level $K$, which can be accomplished by the method outlined in Section~\ref{sec:counts}. The states fall into two blocks, descendants of the vacuum and of an $\mathfrak{so}(N^2-1)$ vector formed by acting with the lowest Fourier mode of $\psi_i$ on the vacuum. The characters are given by \cite{delmastro_infrared_2021}
 \es{chiSO}{
	\chi^{\widehat{\mathfrak{so}}(N^2-1)_1}_{\text{sing}}(q, z) \pm \chi^{\widehat{\mathfrak{so}}(N^2-1)_1}_{\text{vec}}(q, z) = q^{-(N^2 - 1)/48}\prod_{r=1}^\infty \prod_{\lambda \in \adj} \left(1 \pm e^{\lambda \cdot z} q^{r - 1/2}\right)\,.
 }
For example, we have
\begin{equation}\label{eq:so3_characters}
\begin{split}
	q^{1/16}\chi^{\widehat{\mathfrak{so}}(3)_1}_{\text{sing}}(q, z) &= \mathbf{1} + \mathbf{3} q + \left(\mathbf{1} + \mathbf{3} + \mathbf{5}\right) q^2 + \left(\mathbf{1} + 3(\mathbf{3}) + \mathbf{5}\right)q^3 + \left(3(\mathbf{1}) + 4(\mathbf{3}) + 3(\mathbf{5})\right)q^4 + \ldots\,,\\
	q^{1/16}\chi^{\widehat{\mathfrak{so}}(3)_1}_{\text{vec}}(q, z) &= \mathbf{3} q^{1/2} + \left(\mathbf{1} + \mathbf{3}\right)q^{3/2} + \left(\mathbf{1} + 2(\mathbf{3}) + \mathbf{5}\right)q^{5/2} + \left(2(\mathbf{1}) + 3(\mathbf{3}) + 2(\mathbf{5})\right) q^{7/2} + \ldots\,.
\end{split}
\end{equation}

From \eqref{eq:su2_characters} and \eqref{eq:so3_characters}, we see that
\begin{equation}
\begin{split}
	\chi^{\widehat{\mathfrak{so}(3)}_1}_{\text{sing}} = \chi^{\widehat{\mathfrak{su}(2)}_2}_{\mathbf{1}}\,,\qquad
	\chi^{\widehat{\mathfrak{so}(3)}_1}_{\text{vec}} = \chi^{\widehat{\mathfrak{su}(2)}_2}_{\mathbf{3}}\,.
\end{split}
\end{equation}
This is an example of the character decomposition given for general $N$ in Section~\ref{sec:km_list}.

\subsection{Asymptotic State Counts}\label{sec:asymptotics}

Let us now use the characters introduced above to obtain the asymptotic state counts given in \eqref{GotACardy}.  We will do this by first deriving generating functions for the state counts at finite~$N$.   Separately, we will also derive a generating function for the counts in the large $N$ limit. Analyzing the asymptotics of these generating functions demonstrates that the growth is substantially faster in the large $N$ case.

We start with the simplest case, $\SU(2)$. The generating function for all the states is given by \eqref{chiSO}, which in this case reads
 \es{chiSO3}{
	q^{1/16}\left(\chi^{\widehat{\mathfrak{so}}(3)_1}_{\text{sing}}(q, z) + \chi^{\widehat{\mathfrak{so}}(3)_1}_{\text{vec}}(q, z)\right) = \prod_{r=1}^\infty \left(1 + q^{r-1/2}\right)\left(1 + e^z q^{r-1/2}\right) \left(1 + e^{-z} q^{r-1/2}\right) \,.
 }
We are ultimately interested in gauge singlet states; let $A_{2,K}$ be the number of $\SU(2)$ singlet states at level $K$, and let
\begin{equation}
	f_2(q) = \sum_{K = 0}^\infty A_{2,K} q^{K/2} \,.
\end{equation}
Using the fact that every non-trivial integer-spin irrep of $\mathfrak{su}(2)$ has a unique state of charge $1$, the gauge-invariant states can be read off from the character \eqref{chiSO3} by subtracting the coefficients of terms independent of $Z = e^z$ and coefficients of the terms linear in $Z$. Using the Jacobi triple product formula, we can neatly collect all the powers of $Z$:
\begin{equation}
	q^{1/16} \chi^{\widehat{\mathfrak{so}}(3)_1}(q, Z) = \left(\sum_{n\in \mathbb{Z}} q^{n^2/2} Z^n \right)\left(\prod_{r=1}^\infty \frac{1+q^{r-1/2}}{1-q^r}\right) \,.
\end{equation}
It then follows that
\begin{equation}
	f_2(q) = \left(1-q^{1/2}\right) \prod_{r=1}^\infty \frac{1+q^{r-1/2}}{1-q^r}\,.
\end{equation}

We can calculate the growth of the coefficients of this generating function as follows. The infinite product can be rewritten as
\begin{equation}
	\prod_{r=1}^\infty \frac{1+q^{r-1/2}}{1-q^r} = \prod_{m=1}^\infty \frac{1}{\left(1-q^{m/2}\right)^{a_m}}\,,
\end{equation}
with $a_m = 0$ if $m \equiv 2 \pmod{4}$ and $a_m = 1$ otherwise. Thus, the coefficient of $q^{K/2}$ counts integer partitions of $K$ where none of the constituents are congruent to 2 modulo 4. A theorem by Meinardus \cite{actor_infinite_1994,meinardus_asymptotische_1953,meinardus_uber_1954} gives the asymptotic behavior for modified partition problems of this form, in terms of analytic data of the Dirichlet series
\begin{equation}
	D(s) = \sum_{m=1}^\infty \frac{a_m}{m^s} \,.
\end{equation}
In our case, we have $D_2(s) = \left(1 - 2^{-s} + 4^{-s}\right) \zeta(s)$, so all necessary data is simple to compute. For the coefficients of this infinite product, Meinardus' theorem gives the asymptotic behavior $\frac{1}{4\sqrt{2}} K^{-1} \exp\left(\pi \sqrt{K/2}\right)$. We then have to multiply by $(1 - q^{1/2})$, which is equivalent to differentiating with respect to $K$. Thus,
\begin{equation}
	A_{2,K} \sim \frac{\pi}{16 K^{3/2}} \exp\left(\pi \sqrt{\frac{K}{2}}\right) \,,
\end{equation}
where by $f(K) \sim g(K)$ we mean $\lim_{K\to\infty} f(K)/g(K) = 1$.

A similar argument gives generating functions for $A_{N,K}$ with $N\ge 2$. For instance, take $N = 3$. In the character for $\widehat{\mathfrak{so}}(8)_1$, we have three pairs of nonzero weights, which can be written in terms of simple roots as $\pm \alpha_1$, $\pm \alpha_2$, and $\pm (\alpha_1 + \alpha_2)$. Using the Jacobi triple product formula on each pair gives
\begin{equation}
	q^{1/6} \chi^{\widehat{\mathfrak{so}}(8)_1}(q, z) = \left(\sum_{n_1\in \mathbb{Z}} q^{n_1^2/2} e^{n_1 z_1}\right)\left(\sum_{n_2\in \mathbb{Z}} q^{n_2^2/2} e^{n_2 z_2}\right)\left(\sum_{n_3\in \mathbb{Z}} q^{n_3^2/2} e^{n_3(z_1 + z_2)}\right)\left(\prod_{r=1}^\infty \frac{(1 + q^{r-1/2})^2}{(1-q^r)^3}\right).
\end{equation}
From this we need to find the number of singlets. One can show using character orthogonality that in $\SU(3)$ the number of singlet representations appearing in a character can be computed by
\begin{equation}
	n_s^{(3)} = m_{(0,0)} + m_{(1,2)} + m_{(2,1)} - 2m_{(1,1)} - m_{(2,2)}\,,
\end{equation}
where e.g. $m_{(1,2)}$ is the multiplicity of the weight $\alpha_1+2\alpha_2$ in the character. We can read off these multiplicities from the product, and find
\begin{equation}
\begin{split}
	f_3(q) &= \left(\sum_{n\in \mathbb{Z}} \left(q^{3n^2/2} + 2q^{(n+2)^2/2 + (n+1)^2/2 + n^2/2} - 2q^{(n+1)^2 + n^2/2} - q^{(n+2)^2 + n^2/2}\right)\right)\\
	&\quad\times\left(\prod_{r=1}^\infty \frac{(1 + q^{r-1/2})^2}{(1-q^r)^3}\right)\,.
\end{split}
\end{equation}

For $\SU(4)$, the procedure is analogous. We have six pairs of nonzero roots in the product formula for $q^{15/48}\chi^{\widehat{\mathfrak{so}}(15)_1}(q,z)$, and we can apply the Jacobi triple product to each. To extract the number of singlets at each level $K$, we use
\begin{equation}\label{eq:ns4}
\begin{split}
	n_s^{(4)} &= m_{\text{(0, 0, 0)}}-3 m_{\text{(1, 1, 1)}}+m_{\text{(1, 2, 1)}}+2 m_{\text{(1, 2, 2)}}-m_{\text{(1, 2, 3)}}+2 m_{\text{(2, 2, 1)}}-2 m_{\text{(2, 2, 2)}}-2 m_{\text{(2, 3, 2)}} \\
	&\quad+2 m_{\text{(2, 3, 3)}}+m_{\text{(2, 4, 2)}}-m_{\text{(2, 4, 3)}}-m_{\text{(3, 2, 1)}}+2 m_{\text{(3, 3, 2)}}-m_{\text{(3, 3, 3)}}-m_{\text{(3, 4, 2)}}+m_{\text{(3, 4, 3)}}\,.
\end{split}
\end{equation}
The generating function is then
\begin{equation}
	f_4(q) = f_{4,\text{pre}}(q) \times \prod_{r=1}^\infty \frac{\left(1 + q^{r-1/2}\right)^3}{\left(1 - q^r\right)^6} \,,
\end{equation}
where $f_{4,\text{pre}}(q)$ is a prefactor obtained using \eqref{eq:ns4} in a similar manner as for $\SU(3)$.

In general, for $\SU(N)$ the generating function for the number of singlets will take the form
\begin{equation}
	f_N(q) = f_{N,\text{pre}}(q) \times \prod_{r=1}^\infty \frac{\left(1 + q^{r-1/2}\right)^{N-1}}{\left(1 - q^r\right)^{N(N-1)/2}} \,.
\end{equation}
If we apply Meinardus's theorem to the infinite product piece, the relevant Dirichlet series is
\begin{equation}
	D_N(s) = \left(N-1\right)\left(1 + \frac{N-4}{2} 2^{-s} + 2^{-2s}\right) \zeta(s) \,.
\end{equation}
From this we find that the coefficients of the infinite product grow like $\exp\left(\pi\sqrt{\frac{(N^2-1)K}{6}}\right)$. Like in the case of $\SU(2)$, the prefactor should not change the exponential dependence, only the polynomial piece, and so we can conclude
\begin{equation}
	A_{N, K} \sim \exp\left(\pi\sqrt{\frac{(N^2-1)K}{6}}\right)\times \poly(K) \,.
\end{equation}
Noting that $c = \frac{N^2 - 1}{2}$ and $K = 2L_0$, we see that this is consistent with the Cardy formula for the entropy $S = 2\pi \sqrt{\frac{c}{6}\left(L_0 - \frac{c}{24}\right)}$.

We can carry out a similar analysis for the large $N$ state counts, which we denote $A_{\infty, K}$. Here we have to count products of traces of the $B^\dagger$ operators. To illustrate the method, pretend for a moment that these operators were bosonic so that such products could not vanish due to fermionic statistics. The single-trace states are then ``cycles'' of odd numbers of length greater than one. It is well-known that if some combinatorial class $\mathcal{A}$ has generating function $A(z)$, then cycles of that class have the generating function \cite{flajolet_cycle_1991}
\begin{equation}\label{eq:bosonic_st_gf}
	C^\mathcal{A}(z) = \sum_{s=1}^\infty \frac{\phi(s)}{s} \log \frac{1}{1-A(z^s)} \,,
\end{equation}
where $\phi(s)$ is Euler's totient function. We can use this formula with the generating function for odd numbers, $A(z) = \frac{z}{1-z^2}$, keeping in mind that it will also include length-one cycles that must eventually be removed.

To go from the single-trace generating function \eqref{eq:bosonic_st_gf} to the multi-trace generating function, we need to construct multisets of the single-trace states. This is again a well-known problem with a straightforward solution; given a combinatorial class $\mathcal{B}$ with generating function $B(z)$, multisets of $\mathcal{B}$ are counted by
\begin{equation}
	\PE[B(z)] = \exp\left(\sum_{k=1}^\infty \frac{B(z^k)}{k}\right) \,.
\end{equation}
Applying this to $C^\mathcal{A}(z)$, and using $\sum_{s\mid q} \phi(s) = q$, we find
\begin{equation}
	\PE\left\lbrack C^\mathcal{A}(z)\right\rbrack = \prod_{q=1}^\infty \frac{1}{1-A(z^q)} \,.
\end{equation}
Finally, to correct for the fact that we included length-one cycles, we can divide this by $\prod_{\text{odd }n} (1 + z^n)$.

A very similar argument follows for our case where the operators are fermionic. Following the analysis in \cite{flajolet_cycle_1991} but being careful to exclude cycles corresponding to null traces, we find a generating function for fermionic cycles,
\begin{equation}\label{eq:fermionic_st_gf}
	\tilde C^{\mathcal{A}}(z) = \sum_{s=1}^\infty \frac{\phi(s)}{s} \log\frac{1}{1 + (-1)^s A(z^s)} \,.
\end{equation}
Likewise, to form multisets of fermionic objects, we can use a modified plethystic exponential
\begin{equation}
	\widetilde\PE\left\lbrack \tilde B(z)\right\rbrack = \exp\left(\sum_{k=1}^\infty (-1)^{k+1} \frac{\tilde B(z^k)}{k}\right) \,.
\end{equation}
It follows that the generating function $f_\infty(z) = \sum_{K=0}^\infty A_{\infty,K} z^K$ of the large $N$ states is given by
\begin{equation}
    f_\infty(z) = \PE\left\lbrack \frac{\tilde{C}^\mathcal{A}(z) +  \tilde{C}^\mathcal{A}(-z)}{2}\right\rbrack\times\widetilde{\PE}\left\lbrack \frac{ \tilde{C}^\mathcal{A}(z) -  \tilde{C}^\mathcal{A}(-z)}{2}\right\rbrack \prod_{\text{odd n}}\frac{1}{1+z^n} \,,
\end{equation}
where the last factor is again to correct for the one-cycle states appearing in \eqref{eq:fermionic_st_gf}.

Remarkably, when $A(z)$ is odd this product of plethystic exponentials simplifies almost as much as in the bosonic case. Using various identities for the totient function, we find
\begin{equation}
	f_\infty(z) = \prod_{k=1}^\infty \frac{1}{1 + (-1)^k A(z^k)} \prod_{\text{odd }n} \frac{1}{1+z^n}\,.
\end{equation}
For our case with $A(z) = \frac{z}{1-z^2}$, the first factor in this product is
\begin{equation}
	1 + \frac{z}{1 - z - z^2} = 1 + \sum_{k=1}^\infty F_k z^k \,,
\end{equation}
where $F_k$ are the Fibonacci numbers with $F_0 = 0$ and $F_1 = 1$. This factor dominates the asymptotic growth of the coefficients, and so we have
\begin{equation}
	S_{\infty, K} \sim \exp\left(\left(\log \phi\right)K\right) \times \poly(K) \,,
\end{equation}
where $\phi = \frac{1 + \sqrt 5}{2}$ is the golden ratio. In particular, we see that these counts grow like $e^{\alpha_\infty K}$ while the counts for finite $N$ grow like $e^{\alpha_N \sqrt{K}}$. Thus, for any fixed $N$, at large enough $K$ almost all the gauge-invariant states we could write down are null. For instance, at $K = 150$ there are $59,436,131$ physical states for $\SU(2)$ and $3,696,065,286,870,911,100,343,887,617,904$ states for large $N$.

\bibliographystyle{ssg}
\bibliography{main}

\end{document}